\documentclass[a4paper, 11pt]{article}
\usepackage[utf8]{inputenc}
\usepackage{jcappub}
\usepackage{amsfonts}
\usepackage{graphicx}
\usepackage{amssymb}
\usepackage{amsmath}
\usepackage{breqn}
\usepackage{array}
\usepackage{float}
\usepackage{tikz}
\usepackage{subfig}
\usepackage{ulem}
\usetikzlibrary{shapes.geometric, arrows}
\tikzstyle{st} = [rectangle, rounded corners, text width = 3cm, text centered, draw = black ]
\tikzstyle{arrow} = [->,>=stealth]

\newcommand{\Mpl}{M_{_{\rm Pl}}}

\def\sq{\sqrt}

\newcommand{\viz}{\textit{viz.~}}

\title{Viable bounce from non-minimal inflation}
\author{Debottam Nandi, Manjeet Kaur}
\affiliation{Department of Physics and Astrophysics, University of Delhi, Delhi 110007, India}
\emailAdd{dnandi@physics.du.ac.in, mkaur1@physics.du.ac.in}

\begin{document}

\abstract{
	The fundamental difficulty in constructing a viable classical bouncing model is to evade the no-go theorem that states that, simultaneously maintaining the observational bounds on the tensor-to-scalar ratio and the non-Gaussian scalar spectrum is not possible. Furthermore, constructing the bouncing phase leads to numerous instabilities such as gradient, ghost, and so on. Most importantly, the model fails to be an attractor, in general, meaning that the solution heavily depends on the initial conditions, resulting in anisotropic (BKL) instability in the system. In this paper, using conformal transformation, we construct a classical bouncing model from a non-minimal slow-roll inflationary model. As a result of the conformal transformation, we show that the model is free of the above instabilities and that it leads to a smooth transition from bouncing to the traditional reheating scenario. We also look at the dynamical analysis of the system in the presence of a barotropic fluid and discover that there exists a wide range of model parameters that allow the model to avoid the BKL instability, making it a viable alternative to inflationary dynamics.
	
	
}


\maketitle

\section{Introduction}

Inflationary paradigm --- the most successful paradigm of the early Universe \cite{STAROBINSKY198099, Sato:1981, Guth:1981, LINDE1982389, Albrecht-Steinhardt:1982, Linde:1983gd, Mukhanov:1981xt, HAWKING1982295, STAROBINSKY1982175, Guth:1982, VILENKIN1983527, Bardeen:1983, Starobinsky:1979ty}, where the Universe undergoes a brief period of accelerating expansion solves many critical issues including the Horizon problem, and also at the same time, it satisfies all observational constraints \cite{Akrami:2018odb,Aghanim:2018eyx}. However, there still remain some difficulties that inflationary paradigm cannot resolve, e.g., the trans-Planckian problem or the initial singularity problem \cite{Martin:2000xs, Borde:1993xh, Borde:2001nh}, difficulty in ruling out models within the paradigm \cite{Martin:2010hh,Martin:2013tda,Martin:2013nzq,Martin:2014rqa}, etc. These issues lead to in search for alternatives to the inflationary paradigm and the most popular one is the classical non-singular bouncing scenario, where the Universe undergoes a phase of contraction until the scale factor reaches a minimum value before it enters the expanding phase \cite{Novello:2008ra, Cai:2014bea, Battefeld:2014uga, Lilley:2015ksa, Ijjas:2015hcc, Brandenberger:2016vhg}.

However, while bouncing cosmology solves many early Universe issues like the Horizon problem and evades the trans-Planckian or the initial singularity problem, the paradigm itself faces many challenges, arguably even more than the inflationary paradigm. For starters, constructing the bouncing phase itself in this paradigm is extremely difficult as it requires violating null-energy conditions, and as a result, around the bounce, many instabilities such as gradient and ghost appear in the system \cite{Kobayashi:2016xpl, Libanov:2016kfc, Ijjas:2016vtq, Banerjee:2018svi, Cai:2016thi, Cai:2017dyi, Kolevatov:2017voe, Mironov:2018oec, Easson:2011zy, Sawicki:2012pz}. Most importantly, even if one may construct a model evading the instabilities, these models fail to be in line with the observational constraints: a small tensor-to-scalar ratio $(r_{0.002} \lesssim 0.06)$ and simultaneously, a very small scalar non-Gaussianity parameter $\left(f_{\rm NL} \sim \mathcal{O} (1)\right)$ \cite{Cai:2009fn, Gao:2014eaa, Gao:2014hea, Quintin:2015rta, Li:2016xjb, Akama:2019qeh, Kothari:2019yyw} \footnote{This has also been addressed in the past in the context of the so-called pre-big bang scenario based on the string cosmology equations, as reported for instance in a recent review in Ref. \cite{Gasperini:2007vw}.}.

Perhaps the most subtle yet important aspect of the early universe is the initial conditions themselves. Gravity being highly non-linear in nature brings the obvious question of whether a particular solution, \viz the evolution of the early Universe depends on the initial condition. If it is independent, i.e., the solution is an attractor, then even if we choose a slight deviation in the initial condition, the Universe will always converge on the desired evolution. Otherwise, the system may quickly lead to a highly unstable condition. Due to the inherent presence of quantum fluctuations (or any other field(s)), these deviations can always be argued to remain in the system. Moreover, while the current observable Universe is isotropic, the Universe at extremely high energy or near the singularity contains inherent anisotropies that cannot be ignored. The energy corresponding to this anisotropic stress may even grow faster than the energy density responsible for the evolution of the Universe, leading to a highly unstable system. This is referred to as the Belinsky-Khalatnikov-Lifshitz (BKL) instability \cite{doi:10.1080/00018737000101171, Karouby:2010wt, Karouby:2011wj, Bhattacharya:2013ut, Cai:2013vm, Ganguly:2021pke}. For these reasons, the attractor solution is always preferred over any other kind of solution. Unfortunately, while the inflationary solution claims to be an attractor, the majority of the bouncing models do not. However, it is soon realized that one class of bouncing solutions, i.e., ekpyrotic bounce \cite{Levy:2015awa} poses as an attractor and can solve the initial condition problem, arguably even better than the inflationary paradigm. This is because inflation being insensitive to initial conditions is still debatable. For instance, in Ref. \cite{East:2015ggf}, using non-perturbative simulations, the authors showed that the inflationary expansion starts under very specific circumstances. Also, in Ref. \cite{Clough:2016ymm}, the authors pointed out the fact that small field potential fails to start the inflationary expansion under most initial conditions. On the other hand, in Ref. \cite{Garfinkle:2008ei}, authors showed that the ekpyrotic contraction is a `super-smoother', i.e., it is robust to a very wide range of initial conditions and avoids Kasner/mixmaster chaos, and a similar statement could never be proven about any other primordial scenario. However, the simplest model of it fails to be in line with the observation, and this leads to in search for other non-minimal theories such as the Horndeski theories or even beyond the Horndeski theories \cite{Horndeski:1974wa, Gleyzes:2014dya, Kobayashi:2019hrl, Cai:2013vm, Kobayashi:2016xpl, Libanov:2016kfc, Ijjas:2016vtq, Banerjee:2018svi, Cai:2016thi, Cai:2017dyi, Kolevatov:2017voe, Mironov:2018oec, Kobayashi:2016xpl, Quintin:2015rta, Li:2016xjb, Cai:2012va, Ilyas:2020qja, Dobre:2017pnt, Zhu:2021whu}.

It has recently been shown that conformal transformation can play a significant role in constructing such a stable and, at the same time, viable bouncing model of the early Universe \cite{Nandi:2018ooh, Nandi:2019xlj, Nandi:2019xag, Nandi:2019xve}. These works laid the groundwork for the concept that, since most of the slow-roll inflationary solutions satisfy the observations, one can, in principle, conformally transform the model in such a way that the new model leads to a bouncing scale factor solution.  In addition to the scalar field, additional matter such as radiation or stiff matter is kept in the system without performing the conformal transformation. Such a construction implies that the newly bouncing model, as well as the original inflationary models, behave as conformally related during the early stage of the Universe, but the conformal invariance immediately breaks down once the additional matter contributes to the system, \viz reheating scenario. As a result, one can argue that the models, in general, are not conformally connected and the physics carried by themselves differ.  In Refs. \cite{Nandi:2020sif, Nandi:2020szp}, we have shown that the approach indeed leads to a viable bouncing model that averts all the above-mentioned issues, including the BKL instability. However, it has been performed using an inflationary model described in minimal Einstein gravity. In this work, we extend the concept and explore the possibility of constructing a bouncing model using a general non-minimal inflationary model and study all possible consequences.

The article is divided into two parts. The first part focuses on constructing the model by conformally transforming the non-minimal inflationary model, which is given in Sec. \ref{sec:1-bounce-intro}. We then extensively study the observational consequences of the model and show that it indeed satisfies the observation, which is identical to the non-minimal inflationary model. We also show, for the first time, that the bouncing phase arises naturally in the model and it does not lead to any instabilities such as ghost or gradient. Lastly, we find that briefly after the `bounce', the model enters into the oscillatory phase where the conventional reheating era occurs. In this regime, the non-minimal coupling becomes unity, and the bouncing model tends to overlap with the original inflationary model.

In the second part, in Sec. \ref{sec:2-stability}, we focus on the dynamical analysis of the model. In this regard, we consider an additional barotropic fluid\footnote{As mentioned earlier, the fluid is considered to be unaffected by the conformal transformation.}. However, such analysis within the framework of non-minimal gravity is non-trivial, as the coupling to gravity plays a huge role in it. Therefore, the concern further escalates regarding the stability and, even if it is so, the required conditions for the newly constructed bouncing model. 
In this article, we thoroughly investigate these concerns and find that the bouncing model is indeed stable for a wide range of model parameters. Assuming the model must evade the BKL instability, i.e., the instability is occurring due to the additional anisotropic fluid, the upper bound of the model parameter ($\alpha$) turns out to be $\sim 2$. We compare our model to that of the conformal inflationary model and find that, as $\alpha$ approaches zero, i.e., the ekpyrotic bouncing phase, the theory becomes extremely stable, even much more stable than that of the inflationary theory. This essentially establishes that the newly constructed bouncing model can even be preferred over the original inflationary model itself. In Sec. \ref{sec:3-conclu}, we conclude our work with future remarks.

A few words about our conventions and notations are in order at this stage of our discussion. In this work, we work with the natural units such that $\hbar=c=1$, and we define the Planck mass to be $\Mpl \equiv (8\pi G)^{-1/2} = 1$. We adopt the metric signature of $(-, +, +, +)$. Also, we should mention that, while the Greek indices are contracted with the metric tensor $g_{\mu \nu}$, the Latin indices are contracted with the Kronecker delta $\delta_{i j}$. Moreover, we shall denote the partial and the covariant derivatives as $\partial$ and $\nabla$. The overdots and overprimes denote derivatives with respect to the cosmic time~$t$ and the conformal time~$\eta$ associated with the Friedmann-Lema\^{\i}tre-Robertson-Walker (FLRW) line-element, respectively. For the inflationary Universe, we often adopt the e-fold time convention, where $a_I(N)\propto e^{N},$ whereas, for the bouncing Universe, we prefer to use e-N-fold time convention as $a_b(\mathcal{N})\propto e^{\frac{\mathcal{N}^2}{2}},$ where for contraction and expansion, $\mathcal{N}$ is negative and positive, respectively, and $\mathcal{N} = 0$ represents the bounce. Lastly, the sub(super)script `$I$' and $b$ denote the quantity in the inflationary and the bouncing mode, respectively.
\section{Constructing the model}\label{sec:1-bounce-intro}
In this work, we consider the action consisting of a scalar field non-minimally coupled to gravity as
\begin{eqnarray}\label{Eq:non-minimal-action}
	\mathcal{S} = \frac{1}{2} \int {\rm d}^4{\rm \bf x} \sqrt{-g} \left[ f^2(\phi)\, R - g^{\mu \nu}\, \omega(\phi)\, \partial_\mu \phi \partial_\nu \phi - 2\,V(\phi)\right] + S_m (g_{\mu \nu}, \Psi_m),
\end{eqnarray}
where, $g_{\mu \nu}$ is the metric tensor, $R$ is the Ricci scalar, $f(\phi)$ is the non-minimal scalar coupling function, $\omega(\phi)$ is the derivative coupling function of $\phi$, $V(\phi)$ is the scalar potential function, and  $S_m (g_{\mu \nu}, \Psi_m)$ is the action associated with an additional field present in the system. In this article, we assume the fluid to be barotropic in nature, which we will explore in later sections. The corresponding equations of motion can be written as
\begin{eqnarray}
	&& f^2 \left(R_{\mu \nu} -  \frac{1}{2} g_{\mu \nu I} R\right) - 2 \nabla_{\mu} f\, \nabla_{\nu} f - 2 f\,  \nabla_{\mu \nu} f  + 2 g_{\mu \nu I}\,\left( \nabla^\lambda f\, \nabla_\lambda f +  f \,\Box f \right) \nonumber \\
	&& \qquad  =  \left[\omega(\phi) \left(\nabla_{\mu}\phi \nabla_{\nu}\phi - \frac{1}{2} g_{\mu \nu I} \nabla^\lambda \phi \nabla_\lambda \phi \right) - g_{\mu \nu} V + T^{(M)}_{\mu \nu} \right], \\
	&& \Box \phi + \frac{1}{2 \omega(\phi)} \left(\omega_{,\phi}\, \nabla^\lambda\phi \nabla_\lambda\phi - 2 V_{,\phi} + 2 \, f\, f_{,\phi}\, R\right) = 0,
\end{eqnarray} 
where, $A_{, \phi} \equiv \left(\partial A/\partial\phi\right),$ $\Box \equiv g^{\mu \nu}\nabla_{\mu}\nabla_{\nu}$, $R_{\mu \nu}$ is the Ricci tensor, $R \equiv g^{\mu \nu} R_{\mu \nu}$ is the Ricci scalar and $T^{(m)}_{\mu \nu}$ is the energy-momentum tensor associated with the barotropic fluid satisfying the continuity equation
\begin{eqnarray}
	\nabla^{\mu} T^{(m)}_{\mu \nu} = 0.
\end{eqnarray}

\noindent Using the Friedmann-Lema\^{\i}tre-Robertson-Walker (FLRW) line element, describing the homogeneous and isotropic Universe
\begin{eqnarray}\label{Eq:FRWLineElement}
	{\rm d}s^2 = -{\rm d}t^2 + a^2(t)\,{\rm d}{\bf x}^2 = \tilde{a}^2(\eta)\left(-{\rm d}\eta^2 + {\rm d}{\bf x}^2\right), \quad a(t) = \tilde{a}(\eta) \equiv\mbox{scale factor},
\end{eqnarray}
the above equations take the following form:
\begin{eqnarray}
	\label{eq:back00}
	&& 3 f^{2} \, H^2 = - 6 H\,f\,f_{,\phi} \dot{\phi} +  \left(\frac{1}{2} \omega\, \dot{\phi}^2 + V + \rho_{M}\right), \\
	\label{eq:backij}
	&& 2f^2\dot{H}=-2\, (f_{,\phi}^2 + \, f f_{,\phi\phi})\dot{\phi}^2- \, 2f f_{,\phi}\ddot{\phi} +\, 2H\, f f_{,\phi}\dot{\phi} - \, (\omega \dot{\phi}^2 +\, (1+w_m)\rho_m)\\
	\label{eq:eqnfield}
	&& \ddot{\phi} + 3 H \,\dot{\phi} + \frac{1}{2 \omega} \left(\omega_{,\phi} \dot{\phi}^2 + 2V_{,\phi} -2  f f_{,\phi} R \right) = 0, \\
	\label{eq:backmatter}
	&& \dot{\rho}_m + 3 H\left(\rho_m + P_m\right) = 0,
\end{eqnarray}

\noindent where $\rho_m$ and $P_m$ are the energy density and pressure density of the additional fluid, i.e.,
$$T^{(m)}{}^{0}_{0} \equiv -\rho_m, \quad T^{(m)}{}^i_{j} \equiv -P_m\,\delta^{i}_{j}.$$ Since the fluid is barotropic, one can define the equation of state of the fluid as
$$w_m \equiv \frac{P_m}{\rho_m},$$ and it is a constant. For example, if the fluid is dust-matter, the corresponding equation of state is $w_m = 0$ and the energy density of the addition fluid varies as $\rho_m \propto a^{-3}.$ Similarly, $w_m = 1/3$ signifies the additional fluid is radiation-like with $\rho_m \propto a^{-4}.$ Most importantly, stiff anisotropic fluid, which we will explore for the study of BKL instability in the upcoming sections, represents $w_m = 1$ and $\rho_m \propto a^{-6}$.

\subsection{Slow-roll inflationary model}\label{sec:1a-slow-roll}
During the early Universe, the scalar field is assumed to dominate over the barotropic fluid. Therefore, one can ignore the matter containing the barotropic fluid in the action \eqref{Eq:non-minimal-action}. If the scalar field is slowly rolling down the potential $V(\phi)$, the Universe undergoes a quasi-exponential expansion, referred to as the slow-roll inflationary expansion. To understand the dynamics, let us define the slow-roll parameters as
\begin{eqnarray}\label{eq:slow-roll1}
	&&\epsilon^{I}_1=-\frac{\dot{H}_I}{H_I^2},\\
	\label{eq:slow-roll2}
	&&\epsilon^I_2=\frac{\ddot{\phi}}{H_I \dot{\phi}},\\
	\label{eq:slow-roll3}
	&&\epsilon^I_3=\frac{\dot{f}_{I} }{H_I  f_I},\\
	\label{eq:slow-roll4}
	&&\epsilon^I_4=\frac{\dot{E}_I}{2 H_I E_I},
\end{eqnarray}
where $E_I\equiv f_{I}^2(\omega_{I}+6f_{I,\phi}^2)$. The slow-roll conditions are $$\{|\epsilon^I_1|, |\epsilon^I_2|,| \epsilon^I_3|, |\epsilon^I_4|\}\ll1.$$ These conditions ensure the first slow-roll parameter $\epsilon^I_1 \ll 1$, i.e., the scale factor solution is near de-Sitter and remains similar for a sufficient e-folding number. By using these conditions, one can approximate the background equations \eqref{eq:back00} and \eqref{eq:eqnfield} as
\begin{eqnarray}\label{eq:slow-rollH}
	&&H_I\simeq \frac{\sq{V_I}}{\sq{3} f_I},\\
	\label{eq:slow-rollphi}
	&&\dot{\phi}\simeq \frac{(4  f_{I,\phi} V_I -\,  f_{I} V_{I,\phi})}{3\, H\, f_I\,  (6   f_{I,\phi}^2  +\,  \omega_I)}.
\end{eqnarray}
These equations are referred to as the slow-roll equations. Using these equations, one can immediately calculate the first slow-roll parameter as
\begin{eqnarray}\label{eq:slow-roll-epsilon}
	\epsilon^I_1 &\simeq& \frac{(-4 f_{I,\phi} V_{I} +\,  f_{I} V_{I,\phi})(-2 f_{I,\phi} V_{I} +\,  f_{I} V_{I,\phi})}{2\, V_{I}^2\,  (6  f_{I,\phi}^2  +\,  \omega_I)} \nonumber \\
	&\simeq& \frac{(1 - 4 \gamma_I)(1 - 2 \gamma_I)}{2(6\gamma_I^2 + \mu_I^2)},
\end{eqnarray}

\noindent where, $\gamma_I, \mu_I$ are functions of the scalar field and are defined as
\begin{eqnarray}\label{eq:def-gamma-xi}
	\frac{f_{I,\phi} }{f_I }\equiv \gamma_I \frac{ V_{I,\phi}}{ V_I},\quad \frac{\sqrt{\omega_I}}{f_I}\equiv \mu_I \frac{ V_{I,\phi}}{ V_I}.
\end{eqnarray}
Similarly, one can express other slow-roll parameters in terms of $\gamma_I$ and $\mu_I$. As the slow-roll conditions demand all four slow-roll parameters to be extremely smaller than one, one can then immediately show that the required conditions demand the time variation of these new parameters to be extremely small, i.e.,
\begin{eqnarray}\label{eq:gammamuvariation}
	\dot{\gamma_I}/{(H \gamma_I)}\ll 1,\quad \dot{\mu_I}/{H \mu_I} \ll 1,
\end{eqnarray} implying that these parameters are nearly constants, and, most importantly,
\begin{eqnarray}\label{eq:cond_for_slow_roll}
	\gamma_I \simeq \frac{1}{4}^{-},\quad \mbox{and/or} \quad \mu_I \gg \gamma_I.
\end{eqnarray} 

\noindent We exclude the domain $1/4 \leq \gamma_I \leq 1/2$ as these conditions lead to negative $\epsilon^I_1.$ Relations \eqref{eq:gammamuvariation} and \eqref{eq:cond_for_slow_roll}, in order words, ensure that all four slow-roll parameters remain extremely small along with $$\phi_{N} \equiv \frac{\dot{\phi}}{H_I} \ll 1,$$
i.e., the field slowly rolls down the potential. In that case, Eqs. \eqref{eq:slow-rollH} and \eqref{eq:slow-rollphi} then lead to the scale factor solution as a function of $\phi$ as
\begin{eqnarray}\label{Eq:scaleInf}
	a_{I}(\phi) \propto\,\, \exp \left(\int^\phi \mathrm d \phi\,\frac{(6\gamma_I^2 + \mu_I^2)}{\left( 4 \gamma_I -1 \right)} \frac{V_{I, \phi}}{V_I}\right),
\end{eqnarray}
and the conformal time can similarly be expressed as
\begin{eqnarray}\label{eq:conformal-time-inflation}
	\eta (\phi) \propto \exp \left(-\int^\phi \mathrm d \phi\,\frac{(6\gamma_I^2 + \mu_I^2)}{\left( 4 \gamma_I -1 \right)} \frac{V_{I, \phi}}{V_I}\right).
\end{eqnarray}
These solutions are essential in constructing the bouncing model, which we shall use in the later part of this article.


Let us discuss one simple example of a non-minimal slow-roll inflation model which is the most popular and arguably the most successful model of the inflationary Universe: the Higgs inflation model \cite{Bezrukov:2007ep, Bezrukov:2009db}. The action for the Higgs inflation model is 

\begin{eqnarray}\label{eq:Higgs-action}
	\mathcal{S} = \frac{1}{2}\int \mathrm d^4x\,\sqrt{-g}\left\{\left(1 + \xi \phi^2\right) R - \nabla_\mu\phi\nabla^\mu\phi - \frac{\lambda}{2}\phi^4\right\}.
\end{eqnarray}

\noindent In this case, by comparing the above action with the non-minimal action \eqref{Eq:non-minimal-action}, one can define
\begin{eqnarray}\label{eq:HiggsInfFunc}
	f_I(\phi) = \sqrt{1 + \xi \phi^2}, \quad \omega_I(\phi) = 1, \quad V_I(\phi) = \frac{\lambda}{4}\phi^4.
\end{eqnarray}

\noindent Using \eqref{eq:def-gamma-xi}, we then express $\gamma_I$ and $\mu_I$ as functions of $\phi$ as

\begin{eqnarray}\label{eq:gammamu}
	\gamma_I(\phi) = \frac{\xi \phi^2}{4(1 + \xi \phi^2)},\quad \mu_I(\phi) = \frac{\phi}{4\sqrt{1 + \xi\phi^2}},
\end{eqnarray}

\noindent and the relative time variation of them can then be obtained as
\begin{eqnarray}
	\frac{\dot{\gamma_I}}{H_I\gamma_I} \simeq -\frac{8}{\phi^2 + \xi(1 + 6 \xi)\phi^4}, \quad \frac{\dot{\mu_I}}{H_I\mu_I} \simeq -\frac{4}{\phi^2 + \xi(1 + 6 \xi)\phi^4}.
\end{eqnarray}

\noindent Slow-roll conditions require to satisfy relations \eqref{eq:gammamuvariation} and \eqref{eq:cond_for_slow_roll}, which, on order to fulfill these, condition for the scalar field turns out to be $$\xi\phi^2 \gg 1.$$ In simple words, when this condition satisfies, the Universe undergoes the conventional slow-roll dynamics, and the slow-roll equations \eqref{eq:slow-rollH} and \eqref{eq:slow-rollphi} take the following forms
\begin{eqnarray}\label{eq:slowHHiggs}
	H_I &\simeq& \sqrt{\frac{\lambda}{12}}\frac{\phi^2}{\sqrt{1 + \xi\phi^2}}, \\
	\label{eq:slowphidotHiggs}
	\dot{\phi} &\simeq& -\frac{2\sqrt{\lambda}}{\sqrt{3}}\frac{\phi^2\sqrt{1 + \xi \phi^2}}{\phi + \xi (1 + 6 \xi)\phi^3}.
\end{eqnarray}
The model satisfies the observational constraint with $\xi \simeq 47200 \sqrt{\lambda}$ with $\lambda \sim \mathcal{O}(1).$ This implies that, $\xi \simeq 5\times10^4.$ At the CMB pivot scale, $\phi_* \simeq 0.04,$ which defines the scale of inflation. Therefore, at this scale, $\xi \phi_*^2 \simeq 80 \gg 1$, which is consistent with the above analysis. Furthermore, at pivot scale,  $\gamma_I \simeq 0.24, ~\mu_I \simeq 10^{-3},\frac{\dot{\gamma_I}}{H\gamma_I} \simeq -2\times10^{-4}$ and $ \frac{\dot{\mu_I}}{H\mu_I}\simeq -10^{-4}$, implying that $\gamma_I$ is close to $1/4$, and the relative variations of $\gamma_I$ and $\mu_I$, respectively are negligible, reassuring that, at and around the pivot scale, slow-roll dynamics takes place, and the corresponding the slow-roll scale factor solution, by using Eq. \eqref{Eq:scaleInf}, can be obtained as
\begin{eqnarray}
	a_I(\phi) \propto \left(1 + \xi\phi^2\right)^{3/4}\exp{\left(-\frac{(1 + 6\xi)(1 + \xi\phi^2)}{8\xi}\right)}.
\end{eqnarray}
\begin{figure}[h!]
	\centering
	\includegraphics[width=.45\textwidth]{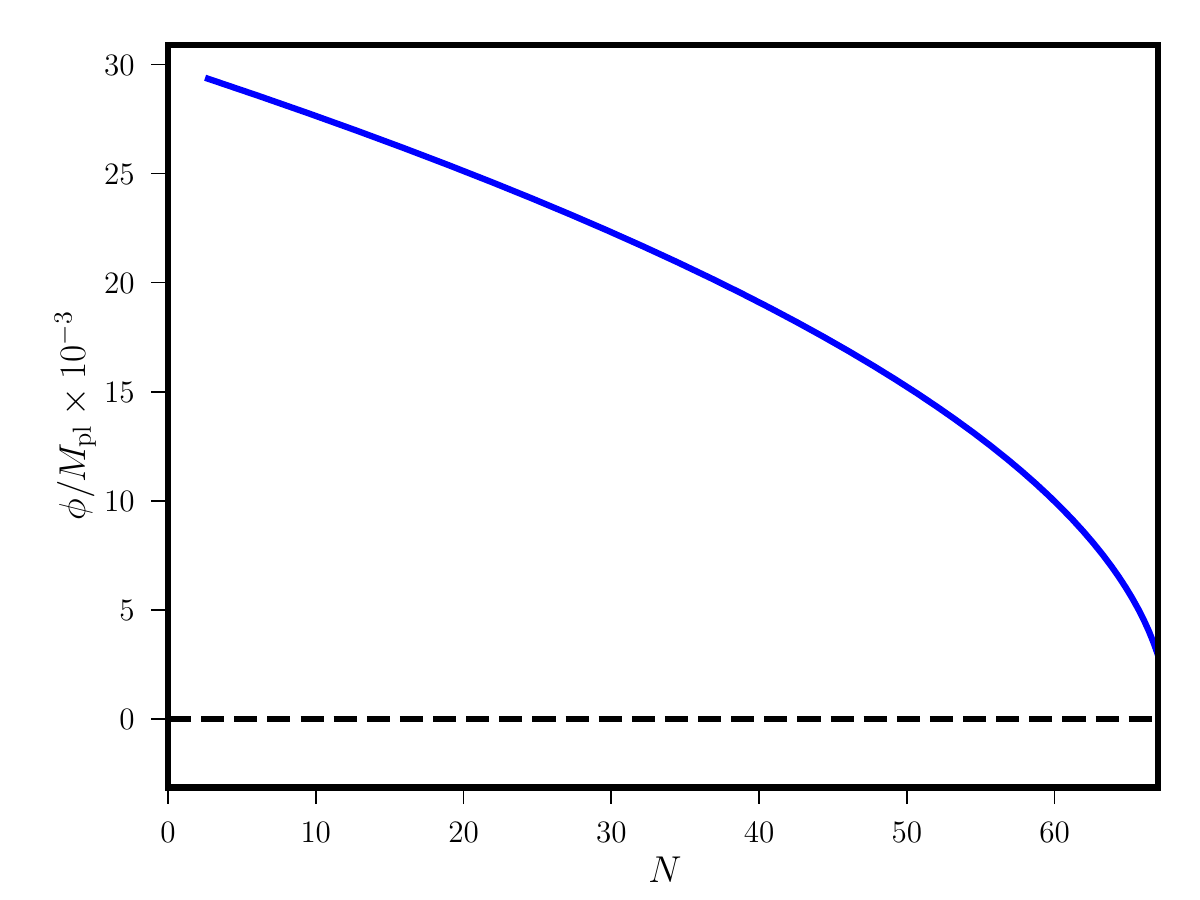}
	\includegraphics[width=.45\textwidth]{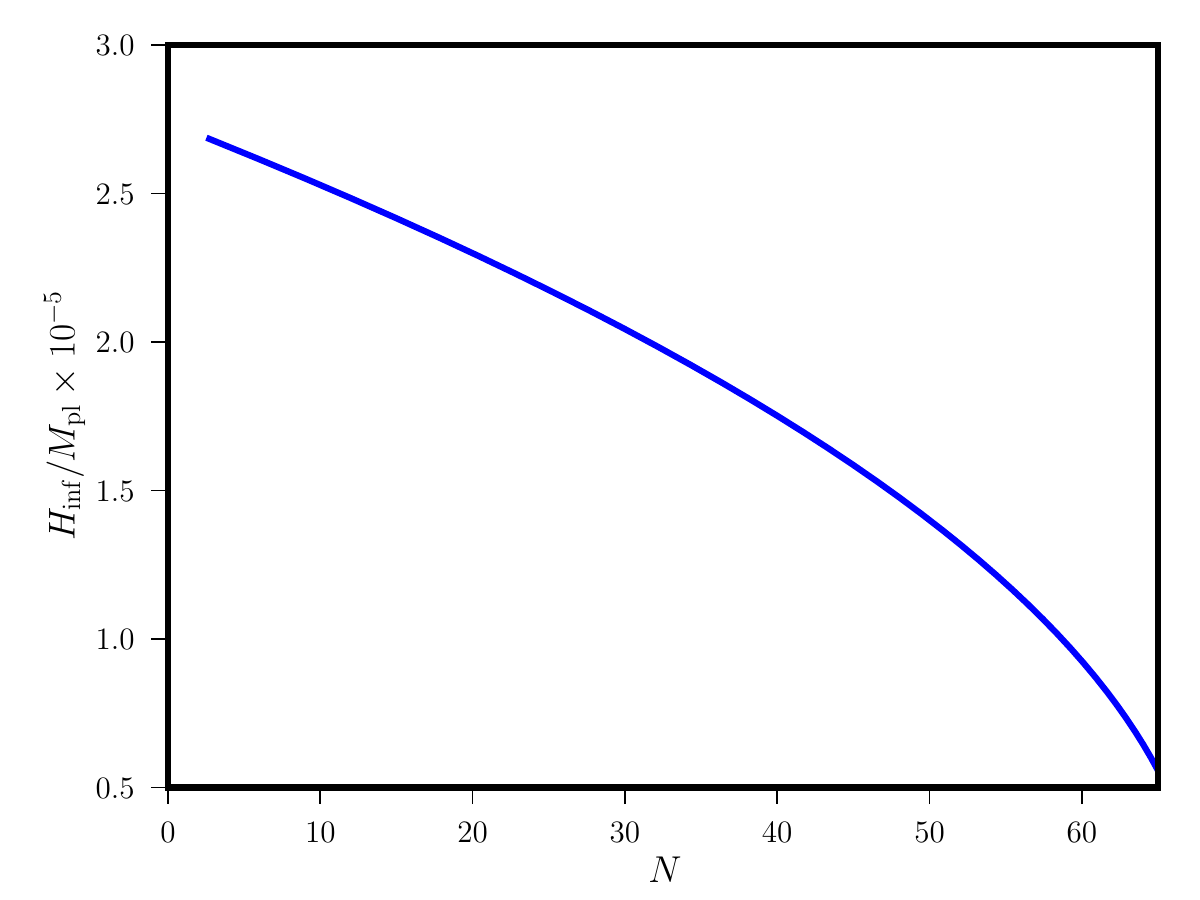}
	\caption{Higgs inflation: we plot the scalar field $\phi$ (left) and the Hubble parameter $H$ (right) as a function of the e-folding number $N$ in the inflationary regime. Note that, the end of inflation occurs at $N = 67.46$ for the choice of initial condition $\phi_i = 0.03, ~\dot{\phi}_i = -10^{-8}$ with $\lambda = 1$ and $\xi = 10^5.$ }
	\label{fig:phi_Hubble_Higgs_inf}
\end{figure}

\begin{figure}[h!]
	\centering
	\includegraphics[width=.47\textwidth]{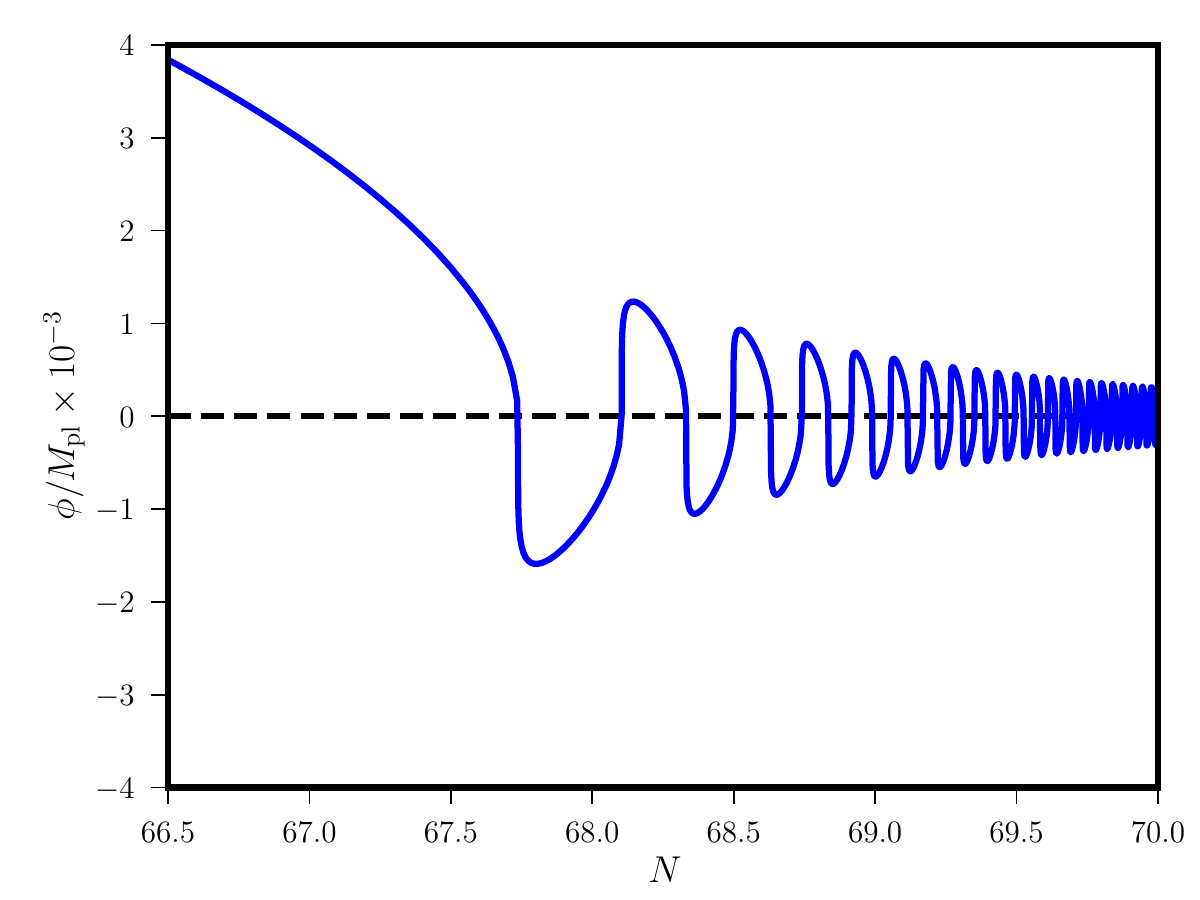}
	\includegraphics[width=.47\textwidth]{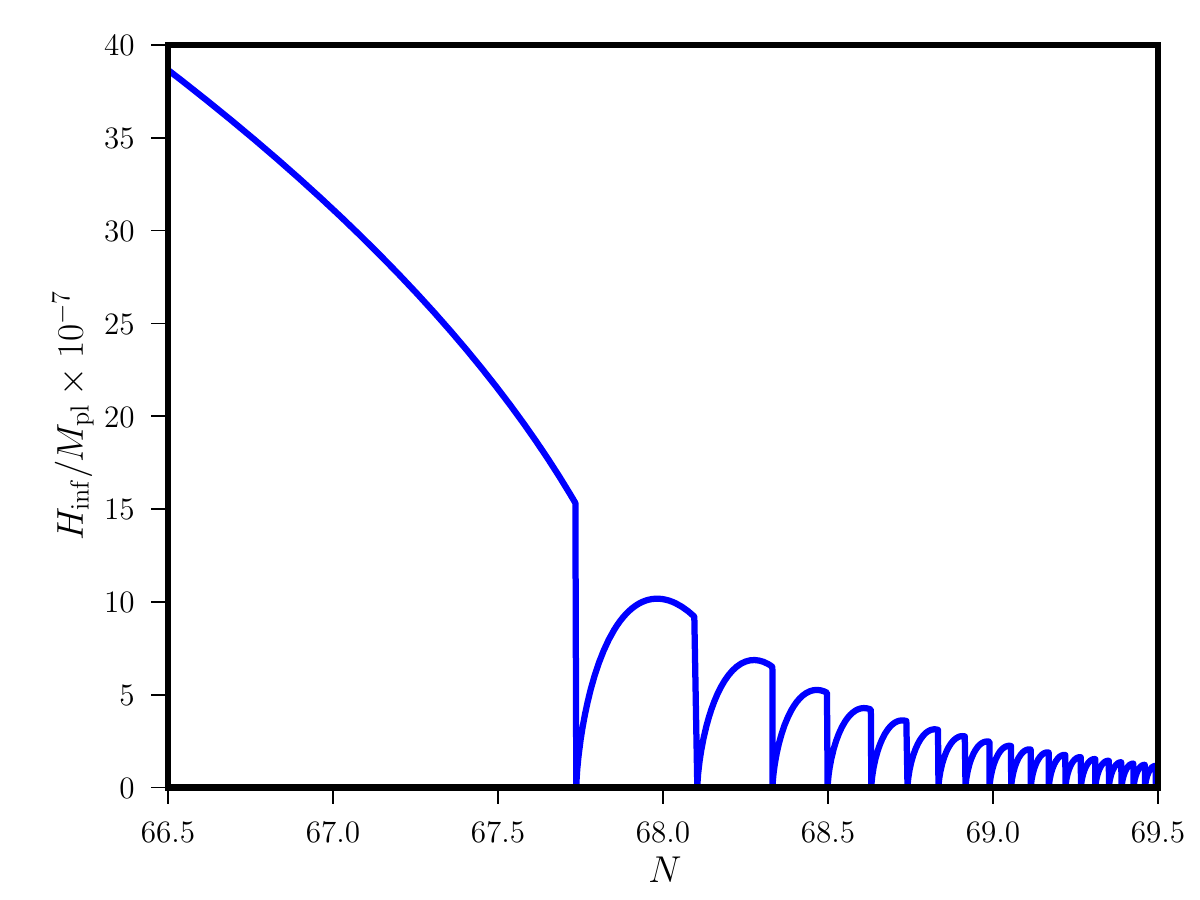}
	\caption{Higgs inflation: we plot the scalar field $\phi$ (left) and the Hubble parameter $H$ (right) as a function of the e-folding number $N$ in the preheating regime. Note that, the end of inflation occurs at $N = 67.46$ for the choice of initial condition $\phi_i = 0.03, ~\dot{\phi}_i = -10^{-8}$ with $\lambda = 1$ and $\xi = 10^5.$ In this regime, the scalar field oscillates around the minimum of the potential and $H$ effectively decays as $a^{-3/2}.$}
	\label{fig:phi_Hubble_Higgs_preheat}
\end{figure}

Please note that the above approximation remains true only at and around the pivot scale and one can extend the approximation till the end of inflation i.e., $\epsilon^I_1 = 1$ with $\phi_e \simeq 5.5 \times 10^{-3}$. After the end of inflation, the slow-roll approximations are violated. At this stage, the field starts oscillating around the minimum of the potential, and the time average value of the Hubble parameter decays as $a^{-3/2}.$ The evolution of the scalar field and the Hubble parameter during and after the end of inflation are shown in Figs. \ref{fig:phi_Hubble_Higgs_inf} and \ref{fig:phi_Hubble_Higgs_preheat}.

\subsection{The bouncing model}

\subsubsection{Background dynamics}

In this section, as mentioned earlier, we construct a model which is conformal to the scalar part of the above inflationary action and we will keep the additional matter part in the original form, as this will ensure the conformal invariance is broken between the two models, and hence the physics. The conformal transformation is a redefinition of the metric field as well as the scale factor as
\begin{eqnarray}\label{Eq:ConfTGen}
	g^{b}_{\mu \nu} = \Omega^2(\phi)\,g^{I}_{\mu \nu} \quad \Rightarrow \quad a_b(\eta) = \Omega(\phi)\,a_I(\eta).
\end{eqnarray}
$g^{b}_{\mu \nu}$ is the new metric describing the bouncing Universe, whereas, $g^{I}_{\mu \nu}$ is the old metric related to the inflationary Universe. Here we are interested in transforming the inflationary scale factor solution into a bouncing solution. Under such transformation, any general non-minimal theory \eqref{Eq:non-minimal-action} transforms into 

\begin{eqnarray}\label{Eq:NonMinAcbounce}
	\mathcal{S}_{b} &=& \frac{1}{2} \int {\rm d}^4{\rm \bf x} \sqrt{-g_{b}} \left[ f_b^2(\phi)\,R_b -\omega_b(\phi) \,g_{b}^{\mu \nu} \partial_\mu \phi \partial_\nu \phi  - 2\, V_{b}(\phi)\right] + S_m.
\end{eqnarray} 
Please note that $S_m$ represents the identical action for the barotropic fluid defined in the original action \eqref{Eq:non-minimal-action} as we are conformally transforming only the scalar sector of the action. The other functions, $f_b(\phi),\, \omega_b(\phi)$ and the potential $V_b(\phi)$  depend on the coupling function $f_I(\phi),~ \omega_{I}(\phi)$ and the inflationary potential $V_I(\phi)$ as
\begin{eqnarray}\label{eq:fomegapotb}
	f_b(\phi)=\frac{f_{I}(\phi)}{\Omega(\phi)},~ \omega_{b}(\phi) = \left(\frac{\omega_{I}}{\Omega^2}+6\frac{f_{I}^2}{\Omega^2}\left(2\frac{f_{I, \phi}}{f_{I}}-\frac{\Omega,_{ \phi}}{\Omega}\right)\frac{\Omega,_{ \phi}}{\Omega}\right),~ V_b(\phi) =\frac{V_I(\phi)}{\Omega(\phi)^4}.
\end{eqnarray}
As the scalar field dominates during the evolution of the early Universe, and assuming that the action \eqref{Eq:non-minimal-action} is responsible for slow-roll inflation, the conformal time can then be expressed as \eqref{eq:conformal-time-inflation}. Under conformal transformation, as the conformal time remains unchanged, we can express the bouncing scale factor solution as $a_b \propto (-\eta)^\alpha, \alpha > 0$ in terms of the scalar field as
\begin{eqnarray}\label{Eq:scaleb}
	a_{b}(\phi) &\propto&\,\, \exp \left(-\alpha\int^\phi \mathrm d \phi\,\frac{(6\gamma_I^2 + \mu_I^2)}{\left(4 \gamma_I - 1 \right)} \frac{V_{I, \phi}}{V_I}\right),
	\quad \alpha > 0,
\end{eqnarray}
which, along with Eq. \eqref{Eq:ConfTGen}, leads to the solution of conformal factor $\Omega(\phi) \propto (-\eta)^{1 + \alpha}$ as

\begin{eqnarray}\label{eq:coupling}
	\Omega(\phi) = \Omega_0\,\exp \left(-(1 + \alpha)\int^\phi \mathrm d \phi\,\frac{(6\gamma_I^2 + \mu_I^2)}{\left(4 \gamma_I - 1 \right)} \frac{V_{I, \phi}}{V_I}\right).
	\quad \alpha > 0.
\end{eqnarray}
$\Omega_0$ is chosen in such a way that, at the bottom of the potential $\Omega(\phi_{\rm min}) = 1$. Please note that $\alpha$, at this level, is still arbitrary, which we shall constrain using the stability analysis in the very next section. Performing the above transformation implies that once the slow-roll inflationary action is chosen, one can obtain the action \eqref{Eq:NonMinAcbounce}, which, along with Eq. \eqref{eq:coupling} leads to the bouncing Universe. This is the main result of this work.

Let us summarize the above: given a non-minimal theory (i.e., $f_I(\phi), \omega_I(\phi), V_I(\phi)$ are known) that leads to slow-roll inflationary dynamics, one can, in principle, construct a bouncing model \eqref{Eq:NonMinAcbounce} with $f_b(\phi), \omega_b(\phi)$ and $V_b(\phi)$ given in \eqref{eq:fomegapotb}. The corresponding bouncing scale factor solution is $a_b(\eta) \propto (-\eta)^\alpha$ and the coupling function $\Omega(\phi)$ in \eqref{eq:coupling} can be expressed in terms of the scalar functions $f_I(\phi), \omega_I(\phi), V_I(\phi).$

Before we proceed any further, let us make a few remarks about the model discussed above: the bouncing solution given in \eqref{Eq:scaleb} holds true only during contraction. In order to understand the evolution of the scale factor in the bouncing model, we have to compare the inflationary and the bouncing Hubble parameter:
\begin{eqnarray}\label{eq:HubbleRelation}
	H_b = \frac{H_I}{\Omega} \left(1 + \frac{\Omega_{,\phi}}{\Omega}\,\phi_{N}\right), \quad \phi_{N} \equiv \frac{\dot{\phi}}{H_I},
\end{eqnarray}
where the overdot and derivative with respect to the e-folding number are performed in the inflationary model. Notice that, with the help of Eqs. \eqref{eq:coupling}, \eqref{eq:slow-rollH} and \eqref{eq:slow-rollphi}, one can show that, during slow-roll evolution, $\frac{\Omega_{,\phi}}{\Omega}\,\phi_{N} \simeq - (1 + \alpha)$. The above equation then leads to $H_b \simeq \alpha\,H_I/\Omega \Rightarrow a_b H_b = -\alpha\, a_I H_I \Rightarrow \mathcal{H}_b = -\alpha \mathcal{H}_I,$ where $\mathcal{H} \equiv \left(a^\prime/a\right).$ During slow-roll, as the scale factor is nearly de-Sitter, i.e., $\mathcal{H}_I \simeq - (1/\eta),$ $\mathcal{H}_b$ turns out to be $(\alpha/\eta)$, which ensures that the corresponding bouncing solution is indeed $a_b(\eta) \propto (-\eta)^\alpha.$ This approximation holds true before the end of inflation, i.e., $\epsilon^I_1 = 1$, as slow-roll approximation is violated thereafter, i.e., $\frac{\Omega_{,\phi}}{\Omega}\,\phi_{N} \neq - (1 + \alpha) $. At this time, as the field approaches the minima of the potential,  $\Omega$ approaches Unity and $\Omega_{,\phi}$ towards zero, indicating that $\frac{\Omega_{,\phi}}{\Omega} \phi_{N}$ smoothly transits from $-(1 + \alpha)$ to zero. It essentially implies that $H_b$ varies from  $-\alpha \frac{H_I}{\Omega}$ to $H_I$, i.e., a negative value to a positive value --- a bouncing phase, and the bounce occurs exactly at $\phi_{N} = -\frac{\Omega}{\Omega_{,\phi}}$ where $H_b = 0.$ After the scalar field reaches the minima of the potential, the field oscillates around the minima, and the conformal factor approaches unity, implying that both the inflationary and the bouncing theory eventually merges with each other. At this time, the conventional reheating scenario occurs where the scalar field couples to the other field (\viz the additional fluid) and decays into other particles. It means the similar (but not \emph{identical}) physics repeats for the bouncing model as well --- reheating after the bounce. Please note that, at this time, in the inflationary model, the dynamics of the Universe become different than the slow-roll dynamics --- implying that dynamics after the bounce are also entirely different than that before the bounce --- making our bouncing model asymmetric.

In order to understand the scenario, let's consider the Higgs inflation model again. In this case, by using Eq. \eqref{eq:HiggsInfFunc} along with Eq. \eqref{eq:coupling}, the conformal coupling can easily be obtained as 
\begin{eqnarray}\label{eq:higgscoupling}
	\Omega(\phi) = \frac{\exp\left({\frac{1}{8} (\alpha +1) (6 \xi +1) \phi ^2}\right)}{ \left(\xi  \phi ^2+1\right)^{\frac{3}{4} (\alpha +1)}}.
\end{eqnarray}
Using the above expression, along with inflationary scalar functions given in \eqref{eq:HiggsInfFunc}, and Eq. \eqref{eq:fomegapotb}, $f_b(\phi), \omega_{b}(\phi)$ and the new bouncing potential $V_b(\phi)$ can be achieved as

\begin{eqnarray}
	f_b(\phi) &=&  \frac{\left(\xi  \phi ^2+1\right)^{\frac{3 \alpha + 5}{4}}}{\exp\left({\frac{1}{8} (\alpha +1) (6 \xi +1) \phi ^2}\right)}, \\
	\omega_{b}(\phi) &=& -\frac{1}{8}  \frac{\left(\xi  \phi ^2+1\right)^{\frac{3 \alpha }{2}+\frac{1}{2}}}{\exp\left({-\frac{1}{4} (\alpha +1) (6 \xi +1) \phi ^2}\right)} \left(3 (\alpha +1)^2 \xi ^2 (6 \xi +1)^2
	\phi ^6+\right. \nonumber\\
	&& \left. 6 (\alpha +1) \xi  (6 \xi +1)  (\alpha -4 \xi +1)\phi ^4 +  \left(3 (\alpha +1)^2-8 (3 \alpha +4) \xi \right)\phi ^2-8\right),\\
	V_b(\phi) &=& \frac{1}{4} \frac{\left(\xi  \phi ^2+1\right)^{3 \alpha +3}}{\exp\left({\frac{1}{2} (\alpha +1) (6 \xi +1) \phi ^2}\right)} \lambda  \phi ^4.
\end{eqnarray}
\noindent Now, since we obtain all essential functions, we can express the Higgs bouncing model whose action is given in \eqref{Eq:NonMinAcbounce}.
\begin{figure}[h!]
	\centering
	\includegraphics[width=.49\textwidth]{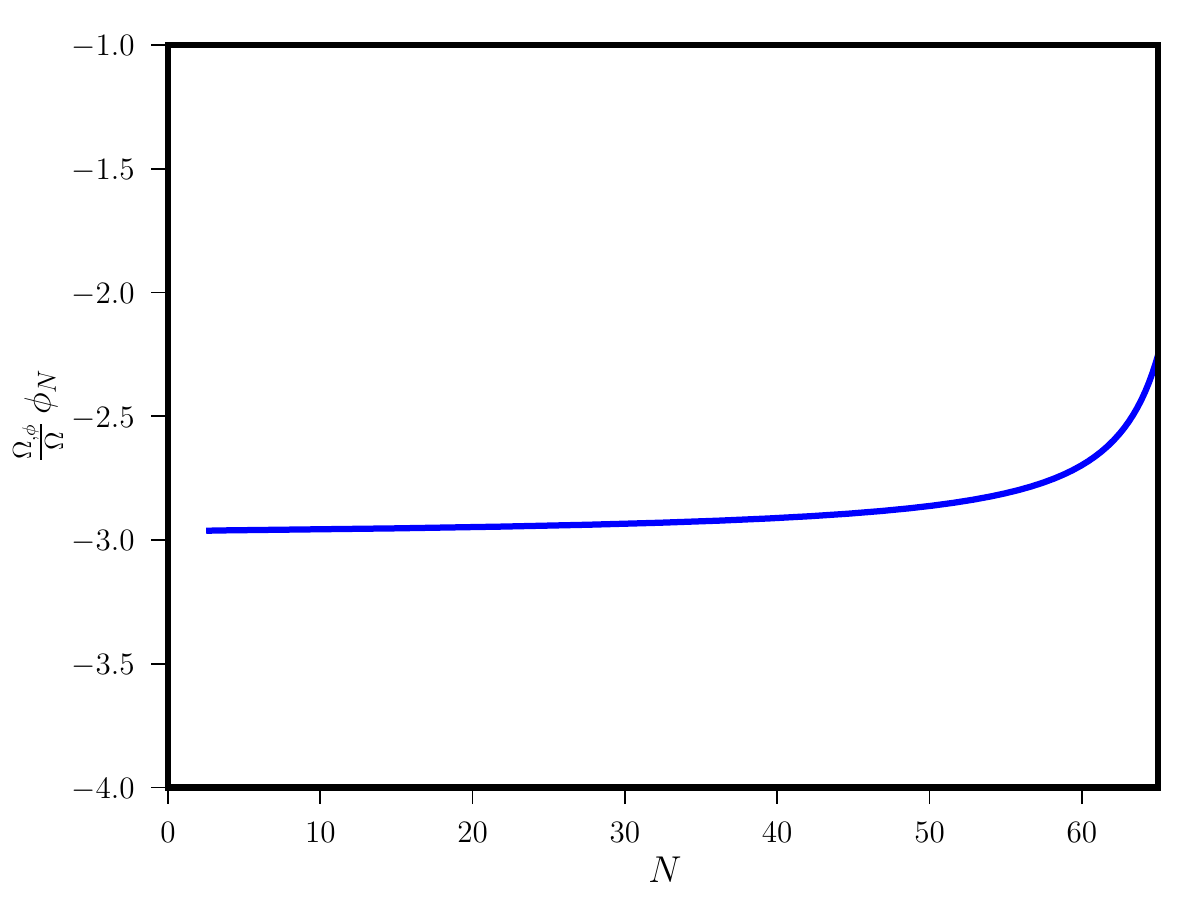}
	\includegraphics[width=.49\textwidth]{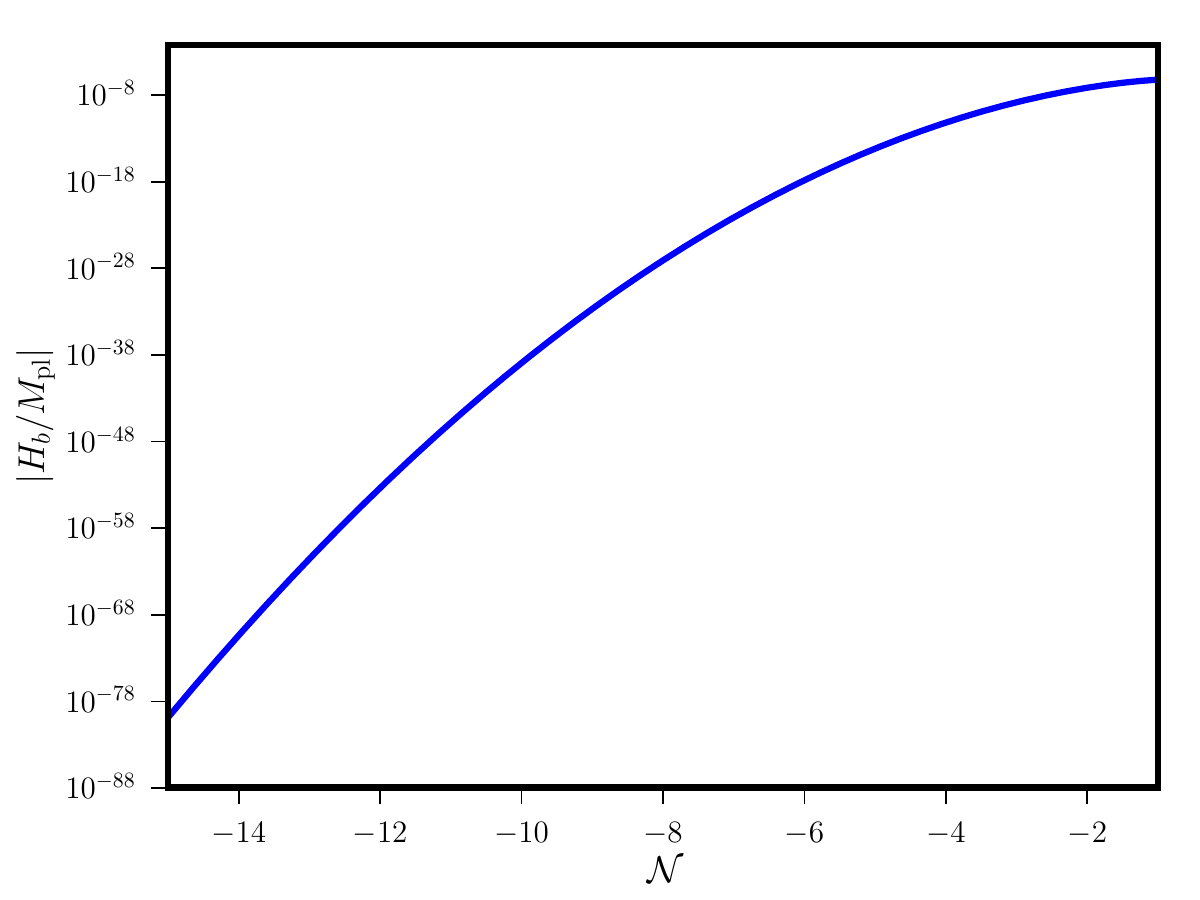}
	\caption{Higgs bounce: on the left, we numerically plot $\frac{\Omega_{,\phi}}{\Omega} \phi_{N}$ as a function of the e-folding number $N$ for $\alpha = 2$. As you can see, it is close to $3$ at the beginning of the inflation, and as it approaches the end of the inflation, the value starts increasing from $-3$. On the right, we numerically plot the bouncing Hubble parameter during the contraction regime as a function of e-N-folding number $\mathcal{N}$. It confirms that the corresponding scale factor solution during contraction is approximately $a_b(\eta) \propto (-\eta)^2.$}
	\label{fig:HiggsContraction}
\end{figure}

\begin{figure}[h!]
	\centering
	\includegraphics[width=.49\textwidth]{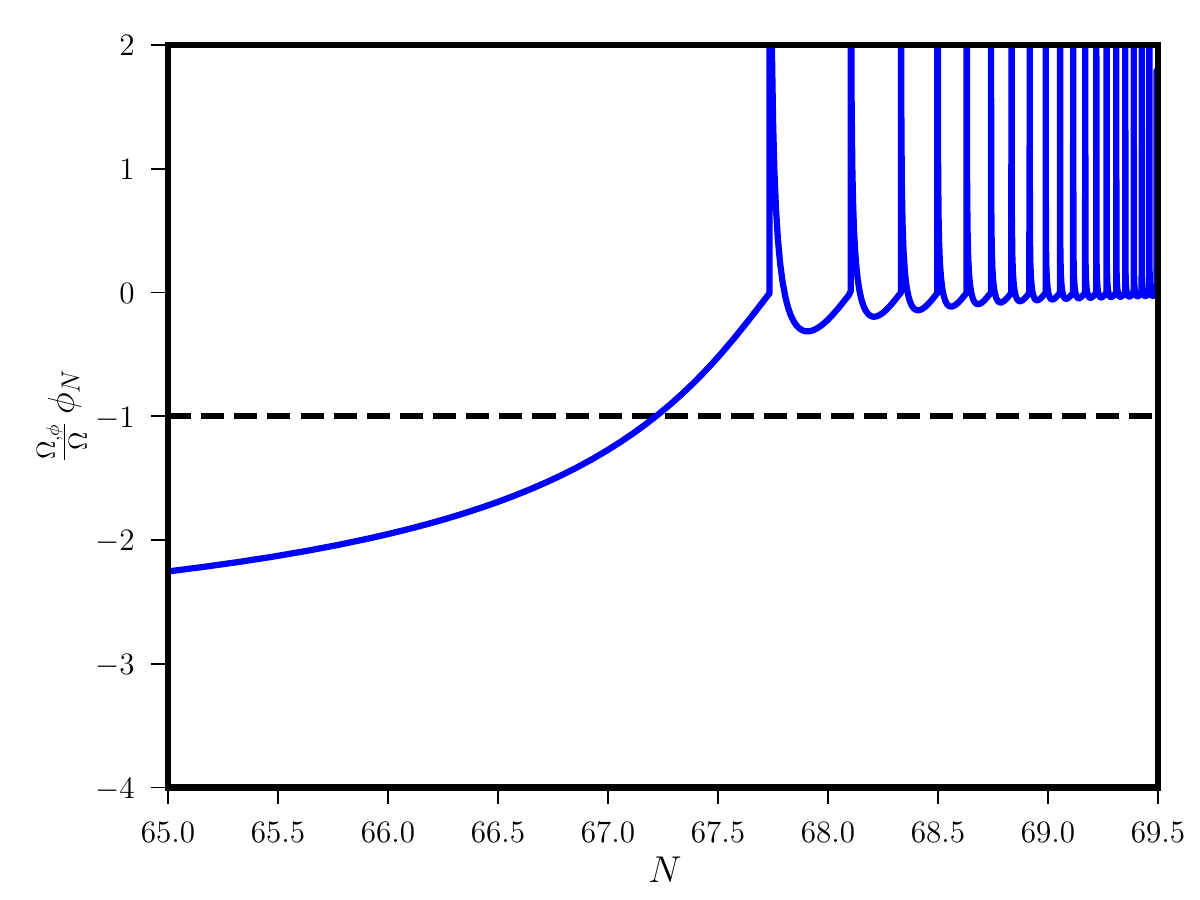}
	\includegraphics[width=.49\textwidth]{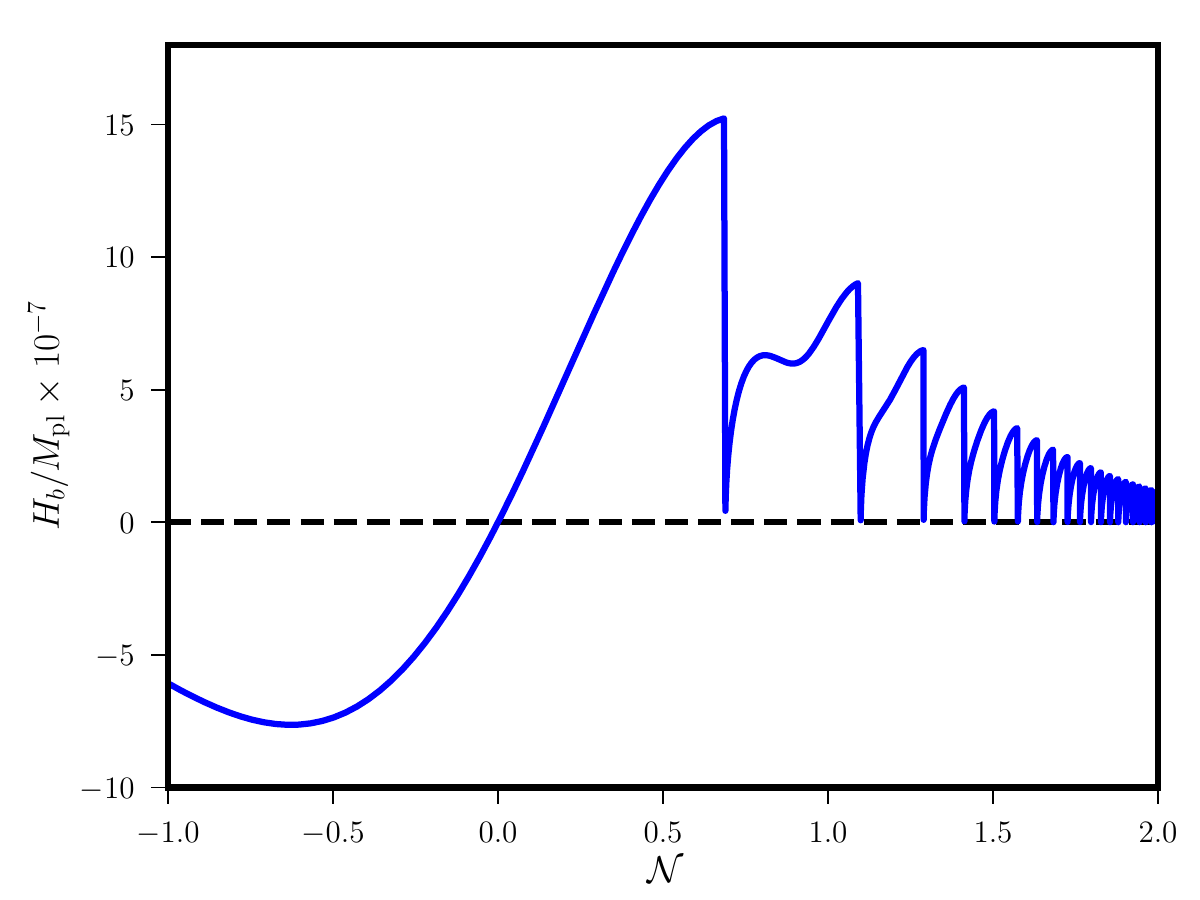}
	\caption{Higgs bounce: on the left, we numerically plot $\frac{\Omega_{,\phi}}{\Omega}\, \phi_{N}$ as a function of e-folding number $N$ for $\alpha = 2$ during and after the end of inflation. As can be seen, during this time, the value gradually increases from the inflationary value of $-3$ to $0.$ On the right, we show that the corresponding Hubble parameter in the bouncing model changes sign from negative to positive, indicating that bounce occurs in the midway.}
	\label{fig:HiggsBounce}
\end{figure}
To understand how the model leads to the dynamics, let us consider the dynamics of the coupling function \eqref{eq:higgscoupling}, most precisely,

\begin{eqnarray}\label{eq:couplingdiff}
	\frac{\Omega_{,\phi}}{\Omega} = \frac{(\alpha +1)  \left( 1 + \xi  (6 \xi +1) \phi ^2\right) \phi}{4(1 +  \xi  \phi ^2)}.
\end{eqnarray}
Using the above equation with Eqs. \eqref{eq:slowHHiggs} and \eqref{eq:slowphidotHiggs}, it becomes obvious that, during slow-roll evolution of the Universe, $1 + \frac{\Omega_{,\phi}}{\Omega} \phi_{N} \simeq -\alpha.$ It tells us that the Hubble parameter is negative and the corresponding scale factor solution is $a_b(\eta) \propto (-\eta)^\alpha$: a contracting phase, which is shown in Fig. \ref{fig:HiggsContraction} for $\alpha = 2.$ As the scalar field approaches the end of inflation, $\frac{\Omega_{,\phi}}{\Omega} \phi_{N}$ increases from $-(1 + \alpha)$ to zero, as $\phi$ approaches zero (cf. Eq. \eqref{eq:couplingdiff}). This implies that, as stated earlier, the Hubble parameter changes sign from negative to a positive value, i.e., the scale factor smoothly transits from the contraction to the expansion regime and in the midway, at $\frac{\Omega_{,\phi}}{\Omega} \phi_{N} = -1$, the exact bounce occurs. This is demonstrated in Fig. \ref{fig:HiggsBounce}.

\begin{figure}[t]
	\centering
	\includegraphics[width=.49\textwidth]{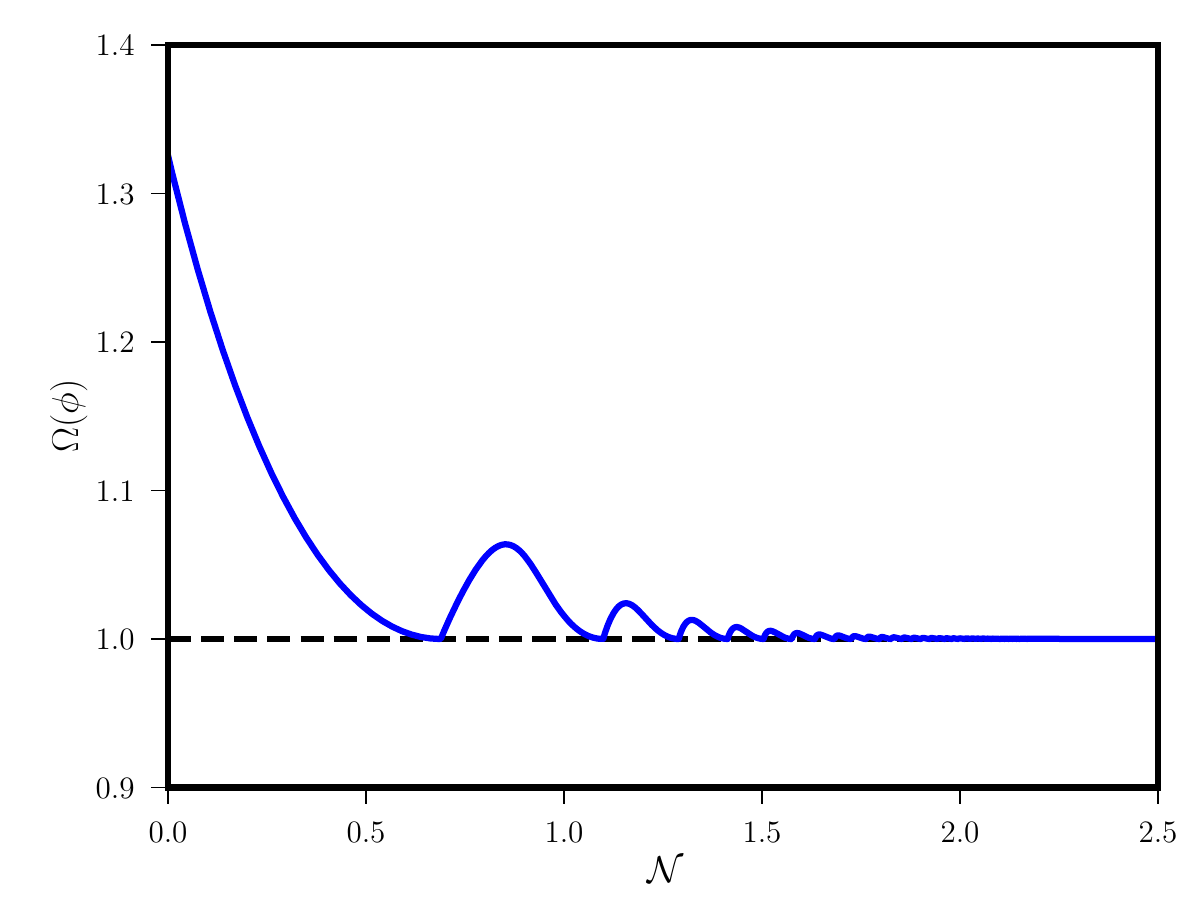}
	\includegraphics[width=.49\textwidth]{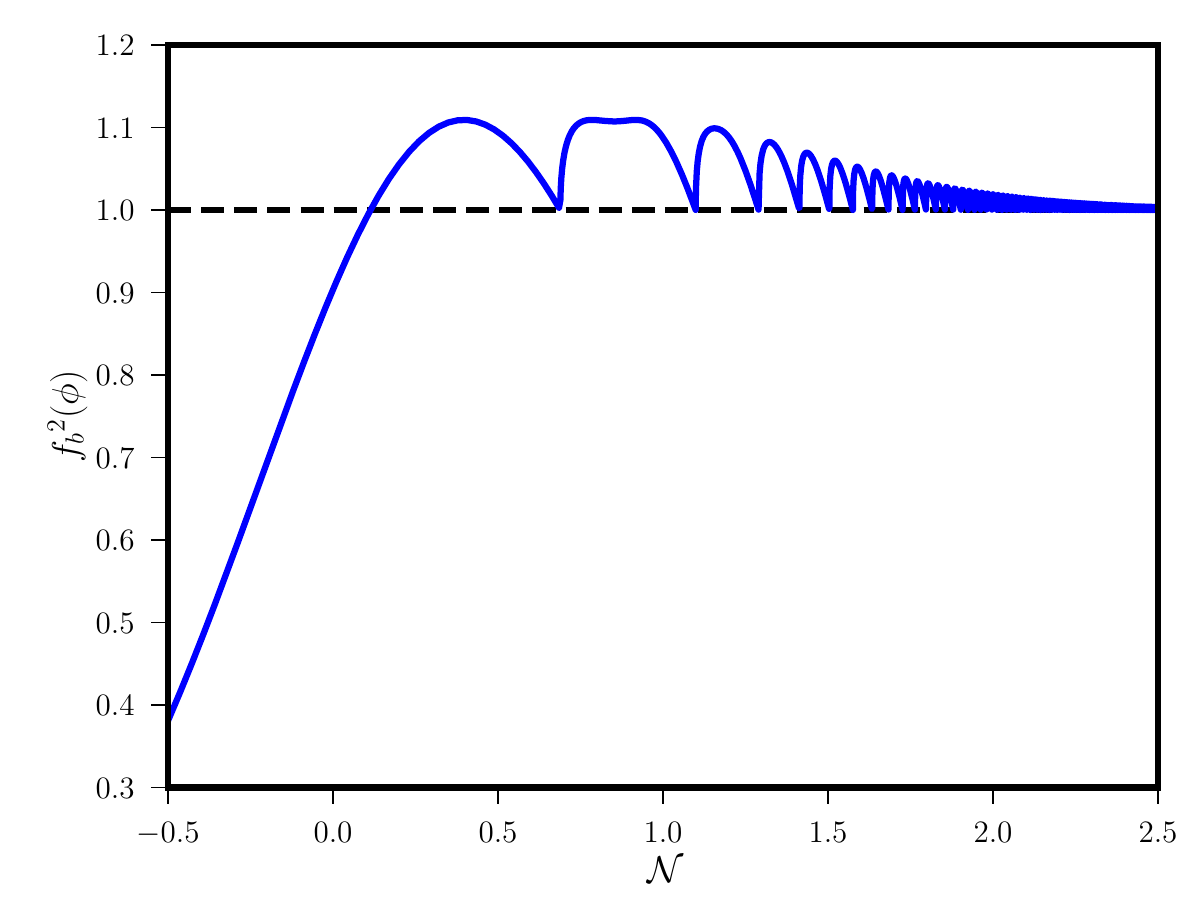}
	\caption{Higgs bounce: we plot the coupling function $\Omega(\phi)$ (left) and $f^2_b(\phi)$ (right) as a function of e-N-fold number $\mathcal{N}$, respectively. It is obvious from the figure that both of them quickly approach unity.}
	\label{fig:couplingfb}
\end{figure}

\begin{figure}[h!]
	\centering
	\includegraphics[width=.9\textwidth]{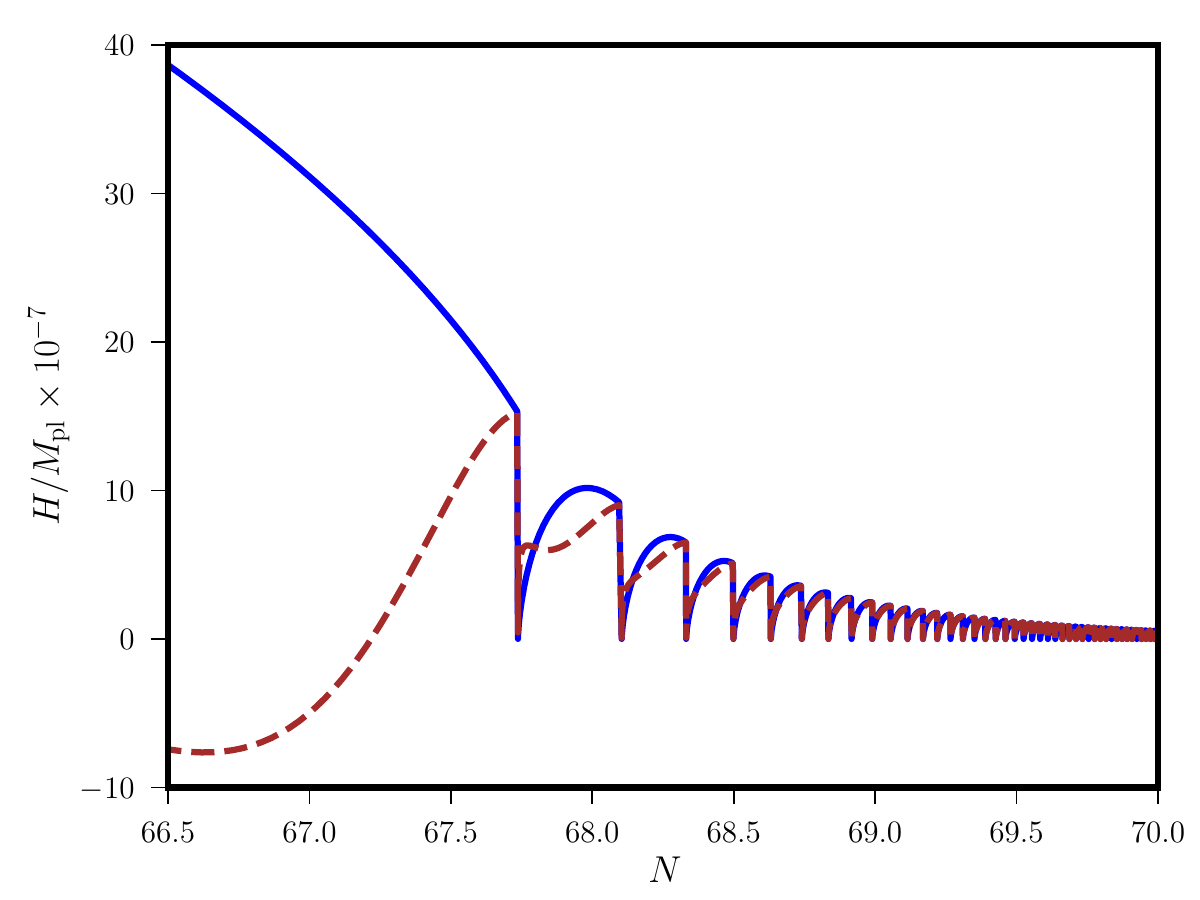}
	\caption{We plot both the inflationary and bouncing Hubble parameters in e-folding number $N$. The one-to-one relationship between the e-fold time $N$ and the e-N-fold time $\mathcal{N}$ convention can be obtained from Eq. \eqref{Eq:ConfTGen}.}
	\label{fig.inf_Bounce_Hubble}
\end{figure}

Shortly after the bounce, as the field $\phi$ reaches the minima of the potential and oscillates around it, the coupling function $\Omega(\phi)$ also oscillates and eventually becomes unity, meaning that, the non-minimal bouncing model merges with the inflationary counterpart (see Fig. \ref{fig:couplingfb}). In fact, the figure tells us that, the function $f_b^2(\phi)$ also manages to approach unity, which implies that, not only both inflationary and bouncing models merge with each other, but also they turn into a minimal gravity theory. The merging of two models is depicted in Fig.  \ref{fig.inf_Bounce_Hubble}, as it shows the Hubble parameter in the bouncing model, as well as the inflationary model, eventually becomes the same (the one-to-one relation between $\mathcal{N}$ and $N$ can be obtained from \eqref{Eq:ConfTGen}). It also tells us that the bounce is asymmetric in nature.

%

\subsubsection{Perturbations}\label{sec:perturb}
As the background evolution of the bouncing model has now been properly examined, let us now concentrate on the scalar and tensor perturbations in the newly constructed model as well. In the case of scalar perturbation, in general, non-minimal gravity \eqref{Eq:non-minimal-action}, the perturbed action is given as 
\begin{eqnarray}
	\delta^2\mathcal{S}_I = \frac{1}{2}\int {\rm d}\eta\, {\rm d \bf x}^3 z_I^2 \left(\zeta_I{}^\prime{}^2 - \left(\nabla \zeta_I\right)^2 \right), \quad z_I(\eta) \equiv  \frac{a_I \phi^\prime}{\mathcal{H}_I}.\frac{\sqrt{\omega_I + 6 f_{I, \phi}^2}}{1 + \frac{f_I^\prime}{\mathcal{H}_I f_I}},
\end{eqnarray}
where $\zeta_I$ is the scalar (also known as the curvature) perturbation and $ \nabla $ is the gradient operator. The corresponding equation of motion is given by

\begin{eqnarray}
	\zeta_I{}^{\prime \prime} + 2 \frac{z_I{}^\prime}{z_I}\zeta_I{}^\prime - \nabla^2\zeta_I = 0.
\end{eqnarray}
The slow-roll dynamics ensure the solution to the above equation leads to freezing of curvature perturbation at the super-Hubble scale and the corresponding spectra behave in a nearly scale-invariant manner. In the non-minimal bouncing Universe with the action \eqref{Eq:NonMinAcbounce}, the action for the scalar perturbation changes by replacing $z_I(\eta)$ with $z_b(\eta)$, i.e.,
\begin{eqnarray}
	\delta^2\mathcal{S}_b = \frac{1}{2}\int {\rm d}\eta\, {\rm d \bf x}^3 z_b^2 \left(\zeta_b{}^\prime{}^2 - \left(\nabla \zeta_b\right)^2 \right), \quad z_b(\eta) \equiv  \frac{a_b \phi^\prime}{\mathcal{H}_b}.\frac{\sqrt{\omega_b + 6 f_{b, \phi}^2}}{1 + \frac{f_b^\prime}{\mathcal{H}_b f_b}}.
\end{eqnarray}
By using Eqs. \eqref{Eq:ConfTGen}, \eqref{eq:fomegapotb} and \eqref{eq:HubbleRelation}, one can immediately show that 
\begin{eqnarray}
	z_b (\eta) = z_I(\eta),
\end{eqnarray}
which, in other words, implies that the conformal time dependence of these two functions is identical. This results in the perturbed bouncing action matching the inflationary model exactly, and since the perturbations are evaluated at the same conformal time, the scalar perturbations in both theories become identical, i.e.,  $\zeta_I = \zeta_b$ at linear order. In a similar way, it is possible to demonstrate that the interaction Hamiltonian (for detailed evaluation, see Refs. \cite{Chen:2006nt, Maldacena:2002vr, Nandi:2015ogk, Nandi:2016pfr, Nandi:2017pfw}) for scalar perturbations at any order remain identical for both inflationary and bouncing models, indicating that the curvature perturbation in the bouncing model at any order is identical to that of inflationary model. This is not surprising because curvature perturbation remains invariant during conformal transformation, as is well-known. Similar to this, the tensor perturbation in the bouncing model remains the same at all orders. Simply put, one can easily create a bouncing model \eqref{Eq:NonMinAcbounce} from an inflationary model (which satisfies all observational constraints) that also produces identical theoretical predictions of observations --- evading the no-go theorem. Also, as in the inflationary model, there is no instability present in the solution of scalar as well as tensor perturbations, the corresponding bouncing model thus contains no instability or divergences (e.g., gradient instability, ghost instability, etc), even at the bounce, meaning that the model leads to stable non-singular bounce.

At this stage, we should note that during the early Universe when the observable modes (i.e., around $k = 0.05$ Mpc${}^{-1}$) leave the Hubble horizon, it is difficult to distinguish between these two models, as the perturbations in two theories act in an identical manner. However, they might not behave similarly during and after the end of inflation, especially during the reheating era. This is because, the two actions, at this level are not conformally the same as the additional matter can play a significant role in the dynamics. It means that the two systems, at this stage, may eventually differ from one another, which in principle may lead to different observational consequences. The analysis of such is beyond the scope of this work and we reserve the work for our future endeavors.

\section{Stability analysis of a non minimally coupled theory}\label{sec:2-stability}
Now that we have already established the bouncing model and have shown that, it is free from all the instabilities and is able to satisfy all the observations, in this section, we shall verify the last subtle yet important issue associated with the bouncing cosmology: the BKL instability. Before we proceed to investigate such issue for the general action, let us make a few remarks. Finding the stability of a generalized system is extremely difficult as it is highly non-linear in nature. However, one can make simple assumptions that may help to simplify the system and as a result, make it solvable around a local area. One such assumption is that the additional fluid is barotropic in nature, which we defined earlier. However, this itself is not enough to solve the system analytically, and therefore, we make another approximation: without the presence of the additional fluid, the corresponding first slow-roll parameter $\epsilon_1$ is `nearly' constant and therefore, the scale factor solution is approximately a power-law in nature. One can always make such approximation when the relative variation of the slow-roll parameter, i.e., $\epsilon_2 \ll 1,$ and as a result, at any given instantaneous time, the scale factor can be approximated as a power law. The corresponding required conditions are $\gamma, \mu$ in Eq. \eqref{eq:def-gamma-xi} to be constants, i.e., 

\begin{eqnarray}\label{eq:gmconst}
\gamma \equiv \frac{V f_{,\phi}}{f V_{,\phi}} = \mbox{Const.},\quad \mu \equiv \frac{V\sqrt{\omega}}{f V_{,\phi}} =\mbox{Const.} 
\end{eqnarray}

\noindent One can then immediately show that, under the above approximation, the (first) slow-roll parameter takes the following exact form as
\begin{eqnarray}
\label{eq:slow-roll-power-law}
\epsilon_1 = -\frac{\dot{H}}{H^2} = \frac{(1-6\gamma+8\gamma^2)}{2(\gamma+2\gamma^2+\mu^2)}.
\end{eqnarray}
Note that, one can compare Eq. \eqref{eq:slow-roll-power-law} with \eqref{eq:slow-roll-epsilon} and show that, at the de-Sitter limit, both coincides with one another. The corresponding scale factor solution can easily be written down as
\begin{eqnarray}
a(t) \propto t^{1/\epsilon_1}.
\end{eqnarray}

\noindent  In the case of minimal gravity,  since $f_{,\phi}$ vanishes, $\gamma$ becomes zero. In this case, the slow-roll parameter and the scale factor solution become

\begin{eqnarray}
\epsilon_1 = \frac{1}{2\mu^2}, \quad a(t) \propto t^{2\mu^2}.
\end{eqnarray}

\noindent Furthermore, in the case of a canonical scalar field minimally coupled gravity, i.e., the simplest model of scalar field theory, $\omega$ is equal to one, which leads to
\begin{eqnarray}
\frac{V_{,\phi}}{V} = \frac{1}{\mu}  = \text{Constant}, \quad \Rightarrow \epsilon_1 = \frac{1}{2}\left(\frac{V_\phi}{V}\right)^2.
\end{eqnarray}
The above result essentially tells us that, in this case, exponential potential leads to the power law scale factor solution, which is well-known in the literature \cite{Lucchin:1984yf}.


However, using Eq. \eqref{eq:gmconst} and the barotropic fluid approximations, one can also show that, along with the power law scale factor solution, it brings other solutions as well, which we will discuss in the next section. Therefore, it is essential to study the stability analysis of our desired power law solution, i.e., whether the solution is independent of the initial conditions (attractor solution). In order to verify the stability, it requires fixed-point analysis of the system \eqref{Eq:non-minimal-action}. In order to do this, it is better to simplify background equations by defining two dimensionless quantities as:
\begin{eqnarray}\label{eq:xy}
x \equiv \sqrt{\frac{\omega}{6}}\frac{\dot{\phi}}{H f}, \quad y \equiv \frac{\sqrt{V}}{\sqrt{3}fH}.
\end{eqnarray}
As mentioned before, since the degrees of freedom of the system is two (each for the scalar field and the barotropic fluid), it is possible to express the evolution of the system only in terms of these dimensionless quantities $x$ and $y$. Using the above definitions of $x$ and $y$ as well as the power-law approximation \eqref{eq:gmconst}, the energy equation \eqref{eq:back00} becomes
\begin{eqnarray}\label{eq:ham_constraint}
\Omega_m \equiv \frac{\rho_m}{3  f^2H^2 } = 1 + 2\, \frac{\sqrt{6}}{\mu} \gamma\, x-y^2 -x^2 ,
\end{eqnarray}
where, $\Omega_M$ is the fractional energy density of the additional fluid. We also find the equations of motion of $x$ and $y$ as
\begin{eqnarray}\label{eq:eomx}
\frac{{\rm d} x}{{\rm d}N} \equiv \frac{1}{H} \frac{{\rm d} x}{{\rm d}t}&=& -\frac{1}{2 \left(6 \gamma ^2 \mu +\mu ^3\right)}\left(-3 x^3 \left(4 \gamma ^2 \mu -\mu ^3 (w_m-1)\right)+\sqrt{6} \gamma  x^2 \left(24 \gamma
	^2+\mu ^2 (7-9 w_m)\right)\right.\nonumber\\
	&&\left.+6 \gamma  \mu  x \left(6 \gamma  w_m+y^2\right)+3 \mu ^3 x
	\left(w_m \left(y^2-1\right)+y^2+1\right)+\sqrt{6} \mu ^2 \left(\gamma  (3 w_m-1)\right.\right.\nonumber\\
	&&\left.\left.+y^2 (1-3
	\gamma  (w_m+1))\right)\right),\\
	\label{eq:eomy}
	\frac{{\rm d} y}{{\rm d}N} \equiv \frac{1}{H} \frac{{\rm d} y}{{\rm d}t} 
	&=& \frac{y}{2 \left(6 \gamma ^2 \mu +\mu ^3\right)} \left(3 x^2 \left(4 \gamma ^2 \mu -\mu ^3 (w_m-1)\right)+\sqrt{6} x \left(6 (1-2 \gamma )
		\gamma ^2+\right.\right.\nonumber\\
		&& \left.\left.\mu ^2 (\gamma  (6 w_m-4)+1)\right)-3 \mu ^3 (w_m+1) \left(y^2-1\right)-6 \gamma
		\mu  \left(y^2-4 \gamma \right)\right),
\end{eqnarray}
where, instead of the cosmic time, we have expressed the time variable as $N$, the e-folding number defined as the change of logarithmic change of scale factor, i.e., $N \equiv \ln{(a)}.$ One can also express the slow-roll parameter in terms of $x$ and $y$ as
\begin{eqnarray}\label{eq:slow-roll-xy}
\epsilon_1 \equiv -\frac{\dot{H}}{H^2} &=&\frac{1}{2 \left(6 \gamma ^2+\mu
	^2\right)}\left(3 \mu ^2 \left(-w_m x^2-(w_m+1) y^2+w_m+x^2+1\right)+2 \sqrt{6} \gamma  \mu  (3
	w_m-1) x\right.\nonumber\\
	&&\qquad\qquad\qquad\left.+6 \gamma  \left(2 \gamma  \left(x^2+2\right)-y^2\right)\right)
\end{eqnarray}
Finally, the effective equation of state in terms of the slow-roll parameter can then be written as
\begin{eqnarray}\label{eq:eeos}
w_{\rm eff} = -1 + \frac{2}{3}\,\epsilon_1,
\end{eqnarray}
which essentially signifies how the effective energy density depends on the scale factor, i.e.,
\begin{eqnarray}
\rho_{\rm eff} \propto a^{-3(1 + w_{\rm eff})}.
\end{eqnarray}


\subsection{Fixed points}\label{sec:fixed-points}\label{sec:1b-fixed-points}
Let us now focus on the model parameters of the non-minimal theory. $\gamma$ and $\mu$ are already two parameters that have been introduced in Eq. \eqref{eq:gmconst}. We also have $w_m$ as the equation of state for the additional barotropic fluid. As a result, the solution and other characteristics can be expressed solely in terms of the three model parameters $\{\gamma, \mu, w_m\}$. Also, by using Eq. \eqref{eq:slow-roll-power-law}, one can express $\mu$ in terms of $\epsilon_1$ and interchangeably use it as a model parameter, i.e., $\{\gamma, \epsilon_1, w_m\}$, which we shall often use in the next section.

Using the evolution equations, one can find the fixed points of the system. These points often represents the solutions of the system, which in this case, describes the dynamics of the Universe depicted by the non-minimal theory \eqref{Eq:non-minimal-action}. These can be found by setting Eqs. \eqref{eq:eomx} and \eqref{eq:eomy} to be equal to zero, i.e., the velocities of $x$ and $y$ vanishes at these points. There are seven such fixed points \cite{Copeland:1997et, Copeland:2006wr, Nandi:2018ooh}:

\begin{eqnarray}
\label{eq:fp1}
&&	1. \quad x^*_1 = \frac{(-1+4\gamma)\mu}{\sqrt{6}(\gamma+\, 2\gamma^2 +\, \mu^2)}, \nonumber\\&& \qquad y^*_1 = \frac{\sqrt{48\gamma^3+120\gamma^4+8\gamma\mu^2+\mu^2(-1+6\mu^2)+\gamma^2(-6+56\mu^2)}}{ \sqrt{6}(\gamma+\, 2\gamma^2 +\, \mu^2)}\\ 
\label{eq:fp2}
&&  2. \quad x^*_2 = \frac{(-1+4\gamma)\mu}{\sqrt{6}(\gamma+\, 2\gamma^2 +\, \mu^2)}, \nonumber\\&& \qquad y^*_2 = -\frac{\sqrt{48\gamma^3+120\gamma^4+8\gamma\mu^2+\mu^2(-1+6\mu^2)+\gamma^2(-6+56\mu^2)}}{ \sqrt{6}(\gamma+\, 2\gamma^2 +\, \mu^2)}\\
\label{eq:fp3}
&&	3. \quad x^*_3 =\frac{\sqrt{6}\gamma -\, \sqrt{6\gamma^2 +\, \mu^2}}{\mu}, \quad y^*_3 = 0 \\
\label{eq:fp4}
&&  4. \quad x^*_4 = \frac{\sqrt{6}\gamma +\, \sqrt{6\gamma^2 +\, \mu^2}}{\mu}, \quad y^*_4 = 0 \\
\label{eq:fp5}
&&	5. \quad x^*_5 =-\sqrt{\frac{3}{2}} (1 + w_m) \mu,\nonumber\\&& \qquad y^*_5 =\frac{\sqrt{\gamma(2 -\, 6w_m) +\, 12(1 +\, w_m)\gamma^2-\, 3 (-1+w_m^2) \mu^2  }}{\sqrt{2}}\\
\label{eq:fp6}
&&  6. \quad x^*_6 =-\sqrt{\frac{3}{2}} (1 + w_m) \mu,\nonumber\\&& \qquad y^*_6 =-\frac{\sqrt{\gamma(2 -\, 6w_m) +\, 12(1 +\, w_m)\gamma^2-\, 3 (-1+w_m^2) \mu^2  }}{\sqrt{2}}\\
\label{eq:fp7}
&&  7. \quad x^*_7 = \sqrt{\frac{2}{3}}\frac{(1 - \, 3\,  w_m)\gamma\mu}{(4\gamma^2 + \, (1 - \, w_m)\mu^2)}, \quad y^*_7 = 0
\end{eqnarray}
Note that, $H$ appears in the denominator in the expression of $y$ in \eqref{eq:xy}. Therefore, while positive sign of $y$ signifies the solution to be expanding, negative sign tells us that the corresponding solution represents contraction. By using Eq. \eqref{eq:slow-roll-xy}, one can then immediately obtain the slow-roll parameter for each of these fixed points and these are
\begin{eqnarray}\label{eq:epsilon-fixed-points}
\epsilon_1(x_1^*, y_1^*) &=& \frac{(1-6\gamma+8\gamma^2)}{2(\gamma+2\gamma^2+\mu^2)}\\
\epsilon_1 (x_2^*, y_2^*) &=& \frac{(1-6\gamma+8\gamma^2)}{2(\gamma+2\gamma^2+\mu^2)}\\
\epsilon_1 (x_3^*, y_3^*) &=& 3+\frac{2\gamma(6\gamma-
	\sqrt{6}\sqrt{6\gamma^2+\mu^2}}{\mu^2}\\
\epsilon_1(x_4^*, y_4^*) &=& 3+\frac{2\gamma(6\gamma+
	\sqrt{6}\sqrt{6\gamma^2+\mu^2}}{\mu^2}\\
\epsilon_1(x_5^*, y_5^*) &=&-\frac{3}{2}(1+w_M)(-1+2\gamma)\\
\epsilon_1(x_6^*, y_6^*) &=&-\frac{3}{2}(1+w_M)(-1+2\gamma)\\
\epsilon_1(x_7^*, y_7^*) &=&\frac{3(-1+w_M)\mu^2-16\gamma^2}{2(-1+w_M)\mu^2-8\gamma^2}.
\end{eqnarray}
Among them, only the first, second, third, and fourth solutions, i.e., Eqs. \eqref{eq:fp1}, \eqref{eq:fp2}, \eqref{eq:fp3}, and \eqref{eq:fp4} are scalar field dominated solutions as the fractional energy density \eqref{eq:ham_constraint} vanishes for each of these solutions, i.e.,
$$\Omega_m (x_1^*, y_1^*) = 0,\quad \Omega_m (x_2^*, y_2^*) = 0, \quad\Omega_m (x_3^*, y_3^*) = 0,\quad \Omega_m (x_4^*, y_4^*)=0.$$
The remaining solutions, Eqs. \eqref{eq:fp5}, \eqref{eq:fp6} and \eqref{eq:fp7} --- refer to the mixed state solutions, in which both the scalar field and the barotropic fluid's energy densities are still non-zero:
\begin{eqnarray}
	&&\Omega_m (x_5^*, y_5^*) = -6 \gamma^2 (w_m+1)-\gamma  (3 w_m+7)-3 \mu ^2 (w_m+1)+1,\nonumber \\
	&&\Omega_m (x_6^*, y_6^*) = -6 \gamma^2 (w_m+1)-\gamma  (3 w_m+7)-3 \mu ^2 (w_m+1)+1,\nonumber\\
	&& \Omega_m (x_7^*, y_7^*) = \frac{\left(6 \gamma ^2+\mu ^2\right) \left(8 \gamma ^2 (2-3 w_m)+3 \mu ^2
		(w_m-1)^2\right)}{3 \left(\mu ^2 (w_m-1)-4 \gamma ^2\right)^2}.
\end{eqnarray}
Note that, in minimal theory with $\gamma = 0$, $\Omega_m (x_5^*, y_5^*)$ and $\Omega_m (x_6^*, y_6^*)$ become one, implying that in that these are fluid dominated fixed points. However, in the non-minimal theory, no such solution exists. Comparing Eq. \eqref{eq:slow-roll-epsilon} with Eqs. \eqref{eq:slow-roll-power-law} and \eqref{eq:epsilon-fixed-points}, and also knowing the solution to be scalar field dominated, i.e., the fractional energy density of the additional fluid to be zero, it is obvious that our desired fixed point is \eqref{eq:fp1} (Eq. \eqref{eq:fp2} is the corresponding contraction solution). We must therefore confirm the solution's stability, which will guarantee that even if we start the evolution away from the desired solution, the initial deviation will quickly go away over time and the deviated solution will always asymptotically merge with the desired solution, and other fixed points won't, in theory, play any part in the evolution.  Therefore, we must assess the Lyapunov exponents that characterize the stability of a fixed point, which are covered in more detail in the following section.

\subsection{Lyapunov exponent and the stability condition}\label{sec:1c-stability}

Now, in order to study the stability of these fixed points, we need to linearize the equations (\ref{eq:eomx}) and (\ref{eq:eomy}) as

\begin{eqnarray}\label{eq:LinearizedEq}
\left(\begin{aligned}
& \frac{{\rm d} \delta x}{{\rm d}N}\\
& \frac{{\rm d} \delta y}{{\rm d}N}
\end{aligned}\right) = \left(\begin{aligned}
\frac{\partial A(x, y)}{\partial x}\Big|_* && \frac{\partial A(x, y)}{\partial y}\Big|_* \\
\frac{\partial B(x, y)}{\partial x}\Big|_* && \frac{\partial B(x, y)}{\partial y}\Big|_*
\end{aligned}\right)\left(\begin{aligned}
\delta x &\\\delta y &
\end{aligned}\right),
\end{eqnarray}
where, $A(x, y)$ and $B(x, y)$ are the right-hand side of (\ref{eq:eomx}) and (\ref{eq:eomy}), respectively. $|_*$ denotes the value at the fixed point. $\delta x$  and $\delta y$ are the deviations of $x$ and $y$ from the corresponding fixed points. By linearizing equations, we presume that we are studying the stability condition close to the fixed point, i.e., the deviation from the fixed point is small. The square matrix mentioned above is needed to be diagonalized in order to determine the eigenvalues and eigenvectors, which in turn will help us find the solutions of $\delta x$ and $\delta y$ as the solutions can be written as

\begin{eqnarray}\label{eq:gen-dev-exp}
&&\delta x = C_{11}\, e^{\lambda_1\, N} + C_{12}\, e^{\lambda_2\,N}, \nonumber \\
&&\delta y = C_{21}\, e^{\lambda_1\, N} + C_{22}\, e^{\lambda_2\,N}.
\end{eqnarray}

\noindent $\lambda_1$ and $\lambda_2$, known as the Lyapunov exponents, are the eigenvalues of the matrix and $C$'s are related to the eigenvectors as well as the initial conditions.  By looking at the above solutions of the deviations, it becomes obvious that, for expanding Universe with $\Delta N > 0,$ i.e., $N$ increases with time, the deviations $\{\delta x, \delta y\}$ decay for both $\{\lambda_1, \lambda_2\} < 0,$ indicating that the fixed point is stable and independent of the initial conditions. This is often referred to as attractor solution of the system. If one or both of the exponents become positive, i.e., the solution is a non-attractor, then $\delta x$ and $\delta y$ grows with time, and as a result, the entire solution quickly moves away from desired fixed point and system may become highly unstable. Similarly, for contracting Universe, the required condition for stability is $\{\lambda_1, \lambda_2\} > 0$ and, at least one of the exponents must be negative for non-attractor solution.

\subsection{Stability of the non-minimal slow-roll inflationary model}\label{sec:non-min-slow-roll-stability}
In this section, we shall evaluate the Lyapunov exponents for slow-roll evolution of the early Universe, i.e., the exponents for the fixed point $\{x_1^*, y_1^*\}$ given in Eq. \eqref{eq:fp1} with $\epsilon_1(x_1^*, y_1^*) \ll 1$, i.e., $\epsilon_1(x_1^*, y_1^*) = \epsilon_1^I$ (cf. \eqref{eq:slow-roll-epsilon}). One can express the Lyapunov exponents in terms of the fixed point and, for the first fixed point \eqref{eq:fp1}, the exponents take the following form:
\begin{eqnarray}\label{eq:LEfp1}
&&\lambda^I_1 = -3 - 3 w_m +\frac{1- 4 \gamma_I}{(\gamma_I +\, 2\gamma_I^2+\, \mu_I^2)},\nonumber\\
&&\lambda^I_2 = - 3 +  \frac{(1 - 4 \gamma_I)(1 + 2 \gamma_I)}{2(\gamma_I +\, 2\gamma_I^2+\, \mu_I^2)}.
\end{eqnarray}

\noindent The above expressions are one of the main result in this work: it establishes the requirement of the stability of the fixed point, i.e., the condition for being an attractor solution. Before examining the stability condition, as mentioned earlier, we find it better to replace the model parameter $\mu_I$ by $\epsilon_1^I$, as we know, for nearly de-Sitter inflation, $\epsilon_1^I \ll 1.$ Using the Eq. \eqref{eq:slow-roll-power-law}, one can invert the relation and by substituting it in the above expression, the two exponents become:
\begin{eqnarray}\label{eq:LyaI}
\lambda_1^I \simeq - 3 - 3 w_m + \frac{1}{(1 - 2 \gamma_I)}\,\epsilon^I_1, \quad \lambda_2^I \simeq - 3 + \left(\frac{1 + 2 \gamma_I}{1 - 2 \gamma_I}\right)\,\epsilon^I_1,
\end{eqnarray}

\noindent where, we use the limit $\epsilon^I_1 \ll 1.$ Therefore, at leading order slow-roll, unless $\gamma_I$ approaches the value $1/2$, the $\lambda$'s remain negative, and can be approximated as $\lambda_{1} \simeq - 3 - 3 w_m,~\lambda_{2} \simeq -3.$ It proves that, slow-roll inflation, i.e., near de-Sitter solution at any non-minimal gravity is an attractor, which is already well known.

Consider the similar example of Higgs inflation model mentioned in the previous section. At the pivot scale, using $\{\gamma_I, \mu_I, w_m\} = \{0.24, 10^{-3}, w_m\},$ the Lyapunov exponents become
\begin{eqnarray}
\lambda_{1}^{\rm Higgs} \simeq -2.9 - 3 w_m, \quad \lambda_{2}^{\rm Higgs} \simeq -2.9,
\end{eqnarray}
which reiterates the fact that, the Higgs inflationary solution is highly stable.

\subsection{Stability of the non-minimal bouncing model}\label{sec:2a-stability-bounce}
Now that we have already established the bouncing model and have shown that it is free from all the instabilities and can satisfy all the observations, in this section, we shall verify the last subtle yet important issue associated with the bouncing cosmology: the BKL instability. Before we proceed to investigate such an issue for general action, let us make a few remarks. Finding the stability of a generalized system is extremely difficult as it is highly non-linear. However, one can make simple assumptions that may help to simplify the system and as a result, make it solvable around a local area. One such assumption is that the additional fluid is barotropic, which we defined earlier. However, this itself is not enough to solve the system analytically, and therefore, we make another approximation: without the presence of the additional fluid, the corresponding first slow-roll parameter $\epsilon_1$ is `nearly' constant, and therefore, the scale factor solution is approximately a power-law in nature. One can always make such an approximation when the relative variation of the slow-roll parameter, i.e., $\epsilon_2 \ll 1,$ and as a result, at any given instantaneous time, the scale factor can be approximated as a power law. The corresponding required conditions are $\gamma, \mu$ in Eq. \eqref{eq:def-gamma-xi} to be constants, i.e., 

\begin{eqnarray}\label{eq:gmconst}
	\gamma \equiv \frac{V f_{,\phi}}{f V_{,\phi}} = \mbox{Const.},\quad \mu \equiv \frac{V\sqrt{\omega}}{f V_{,\phi}} =\mbox{Const.} 
\end{eqnarray}

\noindent One can then immediately show that, under the above approximation, the (first) slow-roll parameter takes the following exact form as
\begin{eqnarray}
	\label{eq:slow-roll-power-law}
	\epsilon_1 = -\frac{\dot{H}}{H^2} = \frac{(1-6\gamma+8\gamma^2)}{2(\gamma+2\gamma^2+\mu^2)}.
\end{eqnarray}
Note that, one can compare Eq. \eqref{eq:slow-roll-power-law} with \eqref{eq:slow-roll-epsilon} and show that, at the de-Sitter limit, both coincide with one another. The corresponding scale factor solution can easily be written down as
\begin{eqnarray}
	a(t) \propto t^{1/\epsilon_1}.
\end{eqnarray}

\noindent  In the case of minimal gravity,  since $f_{,\phi}$ vanishes, $\gamma$ becomes zero. In this case, the slow-roll parameter and the scale factor solution become

\begin{eqnarray}
	\epsilon_1 = \frac{1}{2\mu^2}, \quad a(t) \propto t^{2\mu^2}.
\end{eqnarray}

\noindent Furthermore, in the case of a canonical scalar field minimally coupled gravity, i.e., the simplest model of scalar field theory, $\omega$ is equal to one, which leads to
\begin{eqnarray}
	\frac{V_{,\phi}}{V} = \frac{1}{\mu}  = \text{Constant}, \quad \Rightarrow \epsilon_1 = \frac{1}{2}\left(\frac{V_\phi}{V}\right)^2.
\end{eqnarray}
The above result essentially tells us that, in this case, exponential potential leads to the power law scale factor solution, which is well-known in the literature \cite{Lucchin:1984yf}.


However, using Eq. \eqref{eq:gmconst} and the barotropic fluid approximations, one can also show that, along with the power law scale factor solution, it brings other solutions as well, which we will discuss in the next section. Therefore, it is essential to study the stability analysis of our desired power law solution, i.e., whether the solution is independent of the initial conditions (attractor solution). In order to verify the stability, it requires dynamic analysis of the system \eqref{Eq:non-minimal-action}. In order to do this, it is better to simplify background equations by defining two dimensionless quantities as:
\begin{eqnarray}\label{eq:xy}
	x \equiv \sqrt{\frac{\omega}{6}}\frac{\dot{\phi}}{H f}, \quad y \equiv \frac{\sqrt{V}}{\sqrt{3}fH}.
\end{eqnarray}
As mentioned before, since the degrees of freedom of the system are two (each for the scalar field and the barotropic fluid), it is possible to express the evolution of the system only in terms of these dimensionless quantities $x$ and $y$. Using the above definitions of $x$ and $y$ as well as the power-law approximation \eqref{eq:gmconst}, the energy equation \eqref{eq:back00} becomes
\begin{eqnarray}\label{eq:ham_constraint}
	\Omega_m \equiv \frac{\rho_m}{3  f^2H^2 } = 1 + 2\, \frac{\sqrt{6}}{\mu} \gamma\, x-y^2 -x^2 ,
\end{eqnarray}
where $\Omega_M$ is the fractional energy density of the additional fluid. We also find the equations of motion of $x$ and $y$ as
\begin{eqnarray}\label{eq:eomx}
	\frac{{\rm d} x}{{\rm d}N} \equiv \frac{1}{H} \frac{{\rm d} x}{{\rm d}t}&=& -\frac{1}{2 \left(6 \gamma ^2 \mu +\mu ^3\right)}\left(-3 x^3 \left(4 \gamma ^2 \mu -\mu ^3 (w_m-1)\right)+\sqrt{6} \gamma  x^2 \left(24 \gamma
	^2+\mu ^2 (7-9 w_m)\right)\right.\nonumber\\
	&&\left.+6 \gamma  \mu  x \left(6 \gamma  w_m+y^2\right)+3 \mu ^3 x
	\left(w_m \left(y^2-1\right)+y^2+1\right)+\sqrt{6} \mu ^2 \left(\gamma  (3 w_m-1)\right.\right.\nonumber\\
	&&\left.\left.+y^2 (1-3
	\gamma  (w_m+1))\right)\right),\\
	\label{eq:eomy}
	\frac{{\rm d} y}{{\rm d}N} \equiv \frac{1}{H} \frac{{\rm d} y}{{\rm d}t} 
	&=& \frac{y}{2 \left(6 \gamma ^2 \mu +\mu ^3\right)} \left(3 x^2 \left(4 \gamma ^2 \mu -\mu ^3 (w_m-1)\right)+\sqrt{6} x \left(6 (1-2 \gamma )
	\gamma ^2+\right.\right.\nonumber\\
	&& \left.\left.\mu ^2 (\gamma  (6 w_m-4)+1)\right)-3 \mu ^3 (w_m+1) \left(y^2-1\right)-6 \gamma
	\mu  \left(y^2-4 \gamma \right)\right),
\end{eqnarray}
where, instead of the cosmic time, we have expressed the time variable as $N$, the e-folding number defined as the change of logarithmic change of scale factor, i.e., $N \equiv \ln{(a)}.$ One can also express the slow-roll parameter in terms of $x$ and $y$ as
\begin{eqnarray}\label{eq:slow-roll-xy}
	\epsilon_1 \equiv -\frac{\dot{H}}{H^2} &=&\frac{1}{2 \left(6 \gamma ^2+\mu
		^2\right)}\left(3 \mu ^2 \left(-w_m x^2-(w_m+1) y^2+w_m+x^2+1\right)+2 \sqrt{6} \gamma  \mu  (3
	w_m-1) x\right.\nonumber\\
	&&\qquad\qquad\qquad\left.+6 \gamma  \left(2 \gamma  \left(x^2+2\right)-y^2\right)\right)
\end{eqnarray}
Finally, the effective equation of state in terms of the slow-roll parameter can then be written as
\begin{eqnarray}\label{eq:eeos}
	w_{\rm eff} = -1 + \frac{2}{3}\,\epsilon_1,
\end{eqnarray}
which essentially signifies how the effective energy density depends on the scale factor, i.e.,
\begin{eqnarray}
	\rho_{\rm eff} \propto a^{-3(1 + w_{\rm eff})}.
\end{eqnarray}


\subsection{Fixed points}\label{sec:fixed-points}\label{sec:1b-fixed-points}
Let us now focus on the model parameters of the non-minimal theory. $\gamma$ and $\mu$ are already two parameters that have been introduced in Eq. \eqref{eq:gmconst}. We also have $w_m$ as the equation of state for the additional barotropic fluid. As a result, the solution and other characteristics can be expressed solely in terms of the three model parameters $\{\gamma, \mu, w_m\}$. Also, by using Eq. \eqref{eq:slow-roll-power-law}, one can express $\mu$ in terms of $\epsilon_1$ and interchangeably use it as a model parameter, i.e., $\{\gamma, \epsilon_1, w_m\}$, which we shall often use in the next section.

Using the evolution equations, one can find the fixed points of the system. These points often represent the solutions of the system, which in this case, describes the dynamics of the Universe depicted by the non-minimal theory \eqref{Eq:non-minimal-action}. These can be found by setting Eqs. \eqref{eq:eomx} and \eqref{eq:eomy} to be equal to zero, i.e., the velocities of $x$ and $y$ vanishes at these points. There are seven such fixed points \cite{Copeland:1997et, Copeland:2006wr, Nandi:2018ooh}:

\begin{eqnarray}
	\label{eq:fp1}
	&&	1. \quad x^*_1 = \frac{(-1+4\gamma)\mu}{\sqrt{6}(\gamma+\, 2\gamma^2 +\, \mu^2)}, \nonumber\\&& \qquad y^*_1 = \frac{\sqrt{48\gamma^3+120\gamma^4+8\gamma\mu^2+\mu^2(-1+6\mu^2)+\gamma^2(-6+56\mu^2)}}{ \sqrt{6}(\gamma+\, 2\gamma^2 +\, \mu^2)}\\ 
	\label{eq:fp2}
	&&  2. \quad x^*_2 = \frac{(-1+4\gamma)\mu}{\sqrt{6}(\gamma+\, 2\gamma^2 +\, \mu^2)}, \nonumber\\&& \qquad y^*_2 = -\frac{\sqrt{48\gamma^3+120\gamma^4+8\gamma\mu^2+\mu^2(-1+6\mu^2)+\gamma^2(-6+56\mu^2)}}{ \sqrt{6}(\gamma+\, 2\gamma^2 +\, \mu^2)}\\
	\label{eq:fp3}
	&&	3. \quad x^*_3 =\frac{\sqrt{6}\gamma -\, \sqrt{6\gamma^2 +\, \mu^2}}{\mu}, \quad y^*_3 = 0 \\
	\label{eq:fp4}
	&&  4. \quad x^*_4 = \frac{\sqrt{6}\gamma +\, \sqrt{6\gamma^2 +\, \mu^2}}{\mu}, \quad y^*_4 = 0 \\
	\label{eq:fp5}
	&&	5. \quad x^*_5 =-\sqrt{\frac{3}{2}} (1 + w_m) \mu,\nonumber\\&& \qquad y^*_5 =\frac{\sqrt{\gamma(2 -\, 6w_m) +\, 12(1 +\, w_m)\gamma^2-\, 3 (-1+w_m^2) \mu^2  }}{\sqrt{2}}\\
	\label{eq:fp6}
	&&  6. \quad x^*_6 =-\sqrt{\frac{3}{2}} (1 + w_m) \mu,\nonumber\\&& \qquad y^*_6 =-\frac{\sqrt{\gamma(2 -\, 6w_m) +\, 12(1 +\, w_m)\gamma^2-\, 3 (-1+w_m^2) \mu^2  }}{\sqrt{2}}\\
	\label{eq:fp7}
	&&  7. \quad x^*_7 = \sqrt{\frac{2}{3}}\frac{(1 - \, 3\,  w_m)\gamma\mu}{(4\gamma^2 + \, (1 - \, w_m)\mu^2)}, \quad y^*_7 = 0
\end{eqnarray}
Note that, $H$ appears in the denominator in the expression of $y$ in \eqref{eq:xy}. Therefore, while the positive sign of $y$ signifies the solution to be expanding, a negative sign tells us that the corresponding solution represents contraction. By using Eq. \eqref{eq:slow-roll-xy}, one can then immediately obtain the slow-roll parameter for each of these fixed points and these are
\begin{eqnarray}\label{eq:epsilon-fixed-points}
	\epsilon_1(x_1^*, y_1^*) &=& \frac{(1-6\gamma+8\gamma^2)}{2(\gamma+2\gamma^2+\mu^2)}\\
	\epsilon_1 (x_2^*, y_2^*) &=& \frac{(1-6\gamma+8\gamma^2)}{2(\gamma+2\gamma^2+\mu^2)}\\
	\epsilon_1 (x_3^*, y_3^*) &=& 3+\frac{2\gamma(6\gamma-
		\sqrt{6}\sqrt{6\gamma^2+\mu^2}}{\mu^2}\\
	\epsilon_1(x_4^*, y_4^*) &=& 3+\frac{2\gamma(6\gamma+
		\sqrt{6}\sqrt{6\gamma^2+\mu^2}}{\mu^2}\\
	\epsilon_1(x_5^*, y_5^*) &=&-\frac{3}{2}(1+w_M)(-1+2\gamma)\\
	\epsilon_1(x_6^*, y_6^*) &=&-\frac{3}{2}(1+w_M)(-1+2\gamma)\\
	\epsilon_1(x_7^*, y_7^*) &=&\frac{3(-1+w_M)\mu^2-16\gamma^2}{2(-1+w_M)\mu^2-8\gamma^2}.
\end{eqnarray}
Among them, only the first, second, third, and fourth solutions, i.e., Eqs. \eqref{eq:fp1}, \eqref{eq:fp2}, \eqref{eq:fp3}, and \eqref{eq:fp4} are scalar field dominated solutions as the fractional energy density \eqref{eq:ham_constraint} vanishes for each of these solutions, i.e.,
$$\Omega_m (x_1^*, y_1^*) = 0,\quad \Omega_m (x_2^*, y_2^*) = 0, \quad\Omega_m (x_3^*, y_3^*) = 0,\quad \Omega_m (x_4^*, y_4^*)=0.$$
The remaining solutions, Eqs. \eqref{eq:fp5}, \eqref{eq:fp6} and \eqref{eq:fp7} --- refer to the mixed state solutions, in which both the scalar field and the barotropic fluid's energy densities are still non-zero:
\begin{eqnarray}
	&&\Omega_m (x_5^*, y_5^*) = -6 \gamma^2 (w_m+1)-\gamma  (3 w_m+7)-3 \mu ^2 (w_m+1)+1,\nonumber \\
	&&\Omega_m (x_6^*, y_6^*) = -6 \gamma^2 (w_m+1)-\gamma  (3 w_m+7)-3 \mu ^2 (w_m+1)+1,\nonumber\\
	&& \Omega_m (x_7^*, y_7^*) = \frac{\left(6 \gamma ^2+\mu ^2\right) \left(8 \gamma ^2 (2-3 w_m)+3 \mu ^2
		(w_m-1)^2\right)}{3 \left(\mu ^2 (w_m-1)-4 \gamma ^2\right)^2}.
\end{eqnarray}
Note that, in minimal theory with $\gamma = 0$, $\Omega_m (x_5^*, y_5^*)$ and $\Omega_m (x_6^*, y_6^*)$ become one, implying that in that these are fluid-dominated fixed points. However, in the non-minimal theory, no such solution exists. Comparing Eq. \eqref{eq:slow-roll-epsilon} with Eqs. \eqref{eq:slow-roll-power-law} and \eqref{eq:epsilon-fixed-points}, and also knowing the solution to be scalar field dominated, i.e., the fractional energy density of the additional fluid to be zero, it is obvious that our desired fixed point is \eqref{eq:fp1} (Eq. \eqref{eq:fp2} is the corresponding contraction solution). We must therefore confirm the solution's stability, which will guarantee that even if we start the evolution away from the desired solution, the initial deviation will quickly go away over time and the deviated solution will always asymptotically merge with the desired solution, and other fixed points won't, in theory, play any part in the evolution.  Therefore, we must assess the Lyapunov exponents that characterize the stability of a fixed point, which are covered in more detail in the following section.

\subsection{Lyapunov exponent and the stability condition}\label{sec:1c-stability}

Now, in order to study the stability of these fixed points, we need to linearize the equations (\ref{eq:eomx}) and (\ref{eq:eomy}) as

\begin{eqnarray}\label{eq:LinearizedEq}
	\left(\begin{aligned}
		& \frac{{\rm d} \delta x}{{\rm d}N}\\
		& \frac{{\rm d} \delta y}{{\rm d}N}
	\end{aligned}\right) = \left(\begin{aligned}
		\frac{\partial A(x, y)}{\partial x}\Big|_* && \frac{\partial A(x, y)}{\partial y}\Big|_* \\
		\frac{\partial B(x, y)}{\partial x}\Big|_* && \frac{\partial B(x, y)}{\partial y}\Big|_*
	\end{aligned}\right)\left(\begin{aligned}
		\delta x &\\\delta y &
	\end{aligned}\right),
\end{eqnarray}
where, $A(x, y)$ and $B(x, y)$ are the right-hand side of (\ref{eq:eomx}) and (\ref{eq:eomy}), respectively. $|_*$ denotes the value at the fixed point. $\delta x$  and $\delta y$ are the deviations of $x$ and $y$ from the corresponding fixed points. By linearizing equations, we presume that we are studying the stability condition close to the fixed point, i.e., the deviation from the fixed point is small. The square matrix mentioned above is needed to be diagonalized in order to determine the eigenvalues and eigenvectors, which in turn will help us find the solutions of $\delta x$ and $\delta y$ as the solutions can be written as

\begin{eqnarray}\label{eq:gen-dev-exp}
	&&\delta x = C_{11}\, e^{\lambda_1\, N} + C_{12}\, e^{\lambda_2\,N}, \nonumber \\
	&&\delta y = C_{21}\, e^{\lambda_1\, N} + C_{22}\, e^{\lambda_2\,N}.
\end{eqnarray}

\noindent $\lambda_1$ and $\lambda_2$, known as the Lyapunov exponents, are the eigenvalues of the matrix and $C$'s are related to the eigenvectors as well as the initial conditions.  By looking at the above solutions of the deviations, it becomes obvious that, for expanding Universe with $\Delta N > 0,$ i.e., $N$ increases with time, the deviations $\{\delta x, \delta y\}$ decay for both $\{\lambda_1, \lambda_2\} < 0,$ indicating that the fixed point is stable and independent of the initial conditions. This is often referred to as the attractor solution of the system. If one or both of the exponents become positive, i.e., the solution is a non-attractor, then $\delta x$ and $\delta y$ grow with time, and as a result, the entire solution quickly moves away from desired fixed point and system may become highly unstable. Similarly, for contracting Universe, the required condition for stability is $\{\lambda_1, \lambda_2\} > 0$ and, at least one of the exponents must be negative for the non-attractor solution.

\subsection{Stability of the non-minimal slow-roll inflationary model}\label{sec:non-min-slow-roll-stability}
In this section, we shall evaluate the Lyapunov exponents for slow-roll evolution of the early Universe, i.e., the exponents for the fixed point $\{x_1^*, y_1^*\}$ given in Eq. \eqref{eq:fp1} with $\epsilon_1(x_1^*, y_1^*) \ll 1$, i.e., $\epsilon_1(x_1^*, y_1^*) = \epsilon_1^I$ (cf. \eqref{eq:slow-roll-epsilon}). One can express the Lyapunov exponents in terms of the fixed point and, for the first fixed point \eqref{eq:fp1}, the exponents take the following form:
\begin{eqnarray}\label{eq:LEfp1}
	&&\lambda^I_1 = -3 - 3 w_m +\frac{1- 4 \gamma_I}{(\gamma_I +\, 2\gamma_I^2+\, \mu_I^2)},\nonumber\\
	&&\lambda^I_2 = - 3 +  \frac{(1 - 4 \gamma_I)(1 + 2 \gamma_I)}{2(\gamma_I +\, 2\gamma_I^2+\, \mu_I^2)}.
\end{eqnarray}

\noindent The above expressions are one of the main results in this work: it establishes the requirement of the stability of the fixed point, i.e., the condition for being an attractor solution. Before examining the stability condition, as mentioned earlier, we find it better to replace the model parameter $\mu_I$ by $\epsilon_1^I$, as we know, for nearly de-Sitter inflation, $\epsilon_1^I \ll 1.$ Using the Eq. \eqref{eq:slow-roll-power-law}, one can invert the relation and by substituting it in the above expression, the two exponents become:
\begin{eqnarray}\label{eq:LyaI}
	\lambda_1^I \simeq - 3 - 3 w_m + \frac{1}{(1 - 2 \gamma_I)}\,\epsilon^I_1, \quad \lambda_2^I \simeq - 3 + \left(\frac{1 + 2 \gamma_I}{1 - 2 \gamma_I}\right)\,\epsilon^I_1,
\end{eqnarray}

\noindent where, we use the limit $\epsilon^I_1 \ll 1.$ Therefore, at leading order slow-roll, unless $\gamma_I$ approaches the value $1/2$, the $\lambda$'s remain negative, and can be approximated as $\lambda_{1} \simeq - 3 - 3 w_m,~\lambda_{2} \simeq -3.$ It proves that slow-roll inflation, i.e., near de-Sitter solution at any non-minimal gravity is an attractor, which is already well known.

Consider the similar example of the Higgs inflation model mentioned in the previous section. At the pivot scale, using $\{\gamma_I, \mu_I, w_m\} = \{0.24, 10^{-3}, w_m\},$ the Lyapunov exponents become
\begin{eqnarray}
	\lambda_{1}^{\rm Higgs} \simeq -2.9 - 3 w_m, \quad \lambda_{2}^{\rm Higgs} \simeq -2.9,
\end{eqnarray}
which reiterates the fact that the Higgs inflationary solution is highly stable.

\subsection{Stability of the non-minimal bouncing model}\label{sec:2a-stability-bounce}

We have now reached the last stage of re-evaluating the Lyapunov exponents for the newly built bouncing model. Note that, since the observable modes leave the Hubble horizon during the contracting phase of the bouncing model, we need to ensure the exponents to poses the positive values as the required conditions for stability. In order to accomplish this, we require $\gamma_b$ and $\mu_b$ for the bouncing model. Eq. \eqref{eq:fomegapotb} makes it clear that, $f_{b}(\phi), \omega_b(\phi)$ and $V_{b}(\phi)$ can all be expressed in terms of $f_{I}(\phi), \omega_I(\phi)$ and $V_{I}(\phi)$. As a result, one can express $\gamma_b$ and $\mu_b$ in terms of $\gamma_I$ and $\mu_I$ as
\begin{eqnarray}
	\gamma_b&\equiv& \frac{\gamma_I(-1+2(5+3\alpha)\gamma_I)  +(1+\alpha)\mu_I^2}{(-1 + 4 \gamma_I  + 24\, (1+\alpha)\gamma_I^2 +4(1+ \alpha)\mu_I^2)},\nonumber\\
	\mu_b&\equiv& -\frac{\sqrt{\begin{aligned}
				&(-72\, (1 \, +\, \alpha)\, \gamma_I^3\, (-1\, +\, (7\, +\, 3\alpha)\, \gamma_I)\, -6\, (1\, +\, \alpha)^2\, \mu_I^4 &\\& \qquad \qquad+\, (1\, +\, 4\gamma_I\, (1\, +\, 3\alpha\, -\, 2(13\, +\, 3\alpha(8\, +\, 3\alpha))\, \gamma_I))\, \mu_I^2)&
	\end{aligned}}}{(-1 + 4 \gamma_I  + 24\, (1+\alpha)\gamma_I^2 +4(1+ \alpha)\mu_I^2)  
	}
\end{eqnarray}
The above relations clearly show that, for $\alpha = -1,$ $\gamma_b = \gamma_I,~\mu_b = \mu_I$. This is obvious as, for the same value of $\alpha$, the coupling function in Eq. \eqref{eq:coupling} becomes unity, indicating that the two models should remain entirely unchanged.

Let us first focus on the corresponding fixed point solution. The fixed point in the new conformally modified theory can be obtained by substituting these values in Eq. \eqref{eq:fp1}, where $\gamma_I$ and $\mu_I$ are replaced by $\gamma_b$ and $\mu_b$, respectively:

\begin{eqnarray}
	\label{eq:fp1b1}
	x^*_{b,1} &=& \frac{\sqrt{\begin{aligned}&(-72\, (1 \, +\, \alpha)\, \gamma_I^3\, (-1\, +\, (7\, +\, 3\alpha)\, \gamma_I)\, -6\, (1\, +\, \alpha)^2\, \mu_I^4 \\& \quad \quad \quad +\, (1\, +\, 4\gamma_I\, (1\, +\, 3\alpha\, -\, 2(13\, +\, 3\alpha(8\, +\, 3\alpha))\, \gamma_I))\, \mu_I^2)\end{aligned}}}{\sqrt{6} (\gamma_I(-1 + \gamma_I\, (4+6\alpha))+ \alpha \mu_I^2)}\nonumber\\
	y^*_{b,1} &=& -\frac{\sqrt{48\gamma_I^3+120\gamma_I^4+8\gamma_I\mu_I^2+\mu_I^2(-1+6\mu_I^2)+\gamma_I^2(-6+56\mu_I^2)}}{\sqrt{6}(\gamma_I(-1 + \gamma_I\, (4+6\alpha))+ \alpha\mu_I^2)}
\end{eqnarray}

\noindent As the sign of $y^\ast$ flips to negative, it becomes immediately clear that the universe is indeed contracting. Given how the model was created, the outcome is predictable and self-consistent. Using the slow-roll parameter \eqref{eq:epsilon-fixed-points}, it becomes possible to now observe how the Universe responds to such transformation as

\begin{eqnarray}
	\epsilon^b_1 = \frac{-1+6\gamma_I+4(1+3\alpha)\gamma_I^2+2(1+\alpha)\mu_I^2}{2(-\gamma_I+(4+6\alpha)\gamma_I^2+\alpha\mu_I^2)}.
\end{eqnarray}

\noindent The above expression does not correspond to scale factor solution $a \propto (-\eta)^\alpha.$ However, in order to see the consistency, one must once more express $\mu_I$ in terms of $\epsilon^I_1$ from \eqref{eq:slow-roll-power-law}, and doing so in the above expression leads to

\begin{eqnarray}
	\epsilon_b \simeq \left(1 + \frac{1}{\alpha}\right) + \left(\frac{\alpha (1 - 4 \gamma_I) - 2 \gamma_1}{\alpha^2\,(2 \gamma_I - 1)}\right)\,\epsilon^I_1, \quad \epsilon^I_1\ll 1.
\end{eqnarray}
The above equation establishes the relation between the slow-roll parameters in the newly built bouncing model and the old slow-roll inflationary model. Then the corresponding scale factor solution reduces to

\begin{eqnarray}
	a(\eta) \propto (-\eta)^\beta, \quad \beta \simeq \alpha + \left(\frac{2 \gamma_I + \alpha(4 \gamma_I - 1)}{2 \gamma_I -1}\right)\,\epsilon^I_1.
\end{eqnarray} 
Therefore, unless $\gamma_I$ is close to $1/2$, the bouncing scale factor solution $\beta$ lies very closely to $\alpha.$ To understand this, again consider the Higgs inflation model. In this case, at pivot scale, $$\beta \simeq \alpha - 0.008.$$ Demanding the solution to be contracting at the pivot scale, $\beta$ then requires to be positive and it brings a strong constraint on $\alpha$ as
\begin{eqnarray}
	\alpha_{\rm min} > \frac{2 \gamma_I\epsilon^I_1}{1 - 2 \gamma_I},
\end{eqnarray}
where $\gamma_I$ and $\epsilon^I_1$ are evaluated at the pivot scale.  

In order to determine whether our bouncing solution is stable, let us finally now calculate the Lyapunov exponents for our desired fixed point \eqref{eq:fp1b1}. By following a similar process, the exponents become
\begin{eqnarray}\label{eq:LyaB}
	\lambda^b_1 \simeq
	1 - 3 w_m + \frac{4}{\alpha} + \frac{2(4 \gamma_I (1  + \alpha) - \alpha)}{\alpha^2\,(1 - 2 \gamma_I)}\,\epsilon^I_1,\quad
	\lambda^b_2 \simeq \frac{3}{\alpha} + \frac{(2 \gamma_I (2 \alpha + 3) - \alpha)}{\alpha^2\,(1 - 2\gamma_I)}\,\epsilon^I_1.
\end{eqnarray}
\noindent It is now obvious that, as $\alpha$ approaches zero, i.e., the model is in ekpyrosis, the exponents become extremely high. By using leading order slow-roll approximation, then the two exponents can be approximated as

\begin{eqnarray}\label{eq:lexpb}
	\lambda^b_1 \simeq 1 - 3 w_m  + \frac{4}{\alpha}, \quad \quad\lambda^b_2 \simeq \frac{3}{\alpha},
\end{eqnarray}
where we also assume $\gamma_I$ is not near $1/2.$ Since the Universe is contracting, as obvious from the \eqref{eq:fp1b1}, for the stable contracting solution, we want the above two exponents two be positive. This condition constrains the maximum value of $\alpha$ as

\begin{eqnarray}\label{eq:alphamax}
	\alpha_{\rm max} \simeq \left\{\begin{aligned}
		&\frac{4}{3w_m - 1}, \quad& w_m > 1/3 \nonumber \\
		& \infty,  &w_m \leq 1/3,
	\end{aligned}\right.
\end{eqnarray}

\noindent i.e., when $w_m \leq 1/3,$ the bouncing model is always stable for any given positive value of $\alpha$, and for $w_m \geq 1/3,$ there is an upper bound of $\alpha = \alpha_{\rm max}.$ This is the final result of this work: we transform a slow-roll inflationary model into a bouncing model by using a clever trick of conformal transformation and extensively study the dynamical analysis of the system. We find that the Lyapunov exponents, which characterize the degree of stability, in the newly transformed model are different than the original model (cf. Eq. \eqref{eq:LyaI} and Eq. \eqref{eq:LyaB}), and these exponents highly depend on the model parameters, i.e., $\{\gamma, \mu, w_m\}$ as well as $\alpha.$ Considering the slow-roll parameter in the inflationary model to be sufficiently small, i.e., $\epsilon^I_1 \ll 1,$ the exponents take simpler forms which are given in \eqref{eq:lexpb}. As is clear from these expressions, $\alpha$ approaching zero (i.e., ekpyrotic bounce) results in extremely high values of the exponents which indicates that the solution is extremely stable and can saturate any additional fluid with any $w_m$ value for the equation of state.  The outcome may even demonstrate that the newly developed bouncing model is superior to the original inflationary model in terms of stability, effectively distinguishing itself from its inflationary counterpart.

\subsection{BKL instability and constraints on $\alpha$}\label{sec:2b-BKL-constranit-on-a}
Now let us discuss how the system is affected by the BKL instability. As is well known and discussed before, a system always contains anisotropic stress-energy which corresponds to $\rho_a \propto a^{-6}.$ This essentially tells us that the corresponding equation of state parameter takes the value $w_m = 1,$ which represents stiff matter. In that case, for $\alpha$ and $\gamma_I$ are not close to $0$ and $1/2$, respectively, the two exponents, in this case, take the form
\begin{eqnarray}
	\lambda_1 \simeq -2 + \frac{4}{\alpha}, \qquad \lambda_2 \simeq \frac{3}{\alpha}.
\end{eqnarray}
These conditions lead to the maximum value of the $\alpha$, i.e., $\alpha_{\rm max}$ (cf. Eq. \eqref{eq:alphamax}) and in this case, the stability condition requires that
\begin{eqnarray}
	\alpha_{\rm max} \simeq 2.
\end{eqnarray}
Therefore, the constraint on $\alpha$ is 
\begin{eqnarray}
	\alpha_{\rm min} <\alpha <\alpha_{\rm max} \quad \Rightarrow \frac{2\gamma_I\,\epsilon^I_1}{1 - 2 \gamma_I} <\alpha <2.
\end{eqnarray} Let us try to understand the model with again a realistic example: Higgs inflation. In this case, since $\gamma_I \simeq 0.24$ and $\mu_I \simeq 10^{-3},$ the lower and upper bound on $\alpha$ 
$$\alpha_{\rm \min} \simeq 0.008\quad \alpha_{\rm max} \simeq 2.$$
In these cases, the minimum and maximum values of $\beta$ take the form
\begin{eqnarray}
	\beta_{\rm min} \simeq 0, \quad \beta_{\rm max} \simeq 1.99.
\end{eqnarray}
These values determine the minimum and maximum value of the exponent of the bouncing scale factor solution $a_b(\eta) \propto (-\eta)^\beta$ during contraction, precisely when the pivot scale leaves the horizon. The lowest value of $\alpha$, i.e., $\alpha_{\rm \min}$ causes the maximum stability by enhancing the Lyapunov exponents significantly and the maximum value of these exponents for the Higgs inflation model takes the form 
\begin{eqnarray}
	\lambda_{1}^{\rm Higgs} \simeq 991 - 3 w_m, \quad \lambda_2^{\rm Higgs} \simeq 742,
\end{eqnarray}
which can easily be extended to any other model of slow-roll inflation.

Therefore, similar to other instabilities, we have now established that the bouncing model can also easily evade the BKL instability for a wide range of model parameter $0 \lesssim\alpha \lesssim2.$ This is another main result of this work.

\section{Summary \& conclusions}\label{sec:3-conclu}
In this work, we constructed a new kind of bouncing model with the help of viable slow-roll inflationary models. In doing so, we borrowed the idea of conformal transformation. We conformally transformed the scalar sector of the model and turned the inflationary scale factor solution into a bouncing scale factor solution. In this regard, we took into account the general non-minimal gravity model of inflation and showed that our new scale factor behaves like an asymmetric bouncing solution. Since the curvature and tensor perturbations should be identical due to the conformal relationship between the two models, the newly constructed bouncing models also produce the same spectra as the original inflationary model does, suggesting that the bouncing model can account for the observations. Additionally, just as in the original inflationary model, neither the perturbations nor the bounce cause any instability like ghost, gradient, etc. As a result, the bouncing model is also free of these instabilities. We also demonstrated that, like slow-roll inflation, the model follows an exit mechanism, which leads to the conventional reheating mechanism with the help of an external field. In the end, we also demonstrated that, for $\alpha < 2,$ (model parameter), the model can successfully evade the BKL instability as well, meaning that the model is free from anisotropic instability.

From a model-building perspective, as this is the first attempt to construct a new kind of viable bouncing model, a few issues arise as well. First of all, the issue of the degeneracy of the inflationary paradigm persists here as well, as, by this method of construction, any inflationary model can be converted into a bouncing model. As a result, many inflationary models satisfying the observations lead to a similar number of bouncing models that also satisfy the same constraint. Second, it is seriously questioned whether the bouncing theory can even be observed, as many claims that conformally connected theories are indistinguishable. But because the additional matter content affects the reheating scenarios differently in the two different theories (which has already been mentioned before), this constraint can help distinguish between the inflationary theory and the bouncing theory. Third, one can notice that the $f_b(\phi)$ is extremely small, and as a result, one can think of the model to suffer due to the strong gravity problem. However, this concern has properly been addressed in Refs. \cite{Ageeva:2022asq, Ageeva:2022fyq}.

Let's make a few last points about the range of our upcoming efforts. First, by definition, the parameter $\alpha$ can have any value, including a negative value. This has been discussed before. Second, the conformal transformation was used to construct the new model; however, arbitrary field redefinition, such as disformal transformation, can also be used in a manner that is analogous. Thirdly, and perhaps most significantly, the entire mechanism can be considered a novel method for building models, allowing one to select another formalism, e.g., the Palatini formalism, as well. As a result, the strategy we use in this work not only aids in building bouncing models but also offers a fresh perspective on how to link various theories with various dynamics.

\section*{Acknowledgements}
DN is supported by the DST, Government of India through the DST-INSPIRE Faculty fellowship (04/2020/002142). MK is supported by a DST-INSPIRE Fellowship under the reference number: IF170808, DST, Government of India. DN and MK are also very thankful to the Department of Physics and Astrophysics, University of Delhi. MK and DN also acknowledge facilities provided by the IUCAA Centre for Astronomy Research and Development (ICARD), University of Delhi.


\providecommand{\href}[2]{#2}\begingroup\raggedright\endgroup

\end{document}